\documentclass[
	aps, aps,pra,longbibliography, superscriptaddress, twocolumn,
	10pt
	floatfix, 
    nofootinbib,
	tightenlines
]{revtex4-1}
\usepackage[final]{graphicx}
\usepackage{times,bbm,amsmath,amssymb}
\usepackage{epsfig,color}
\usepackage{xcolor}
\usepackage{hyperref}
\hypersetup{
    colorlinks = true
}
\usepackage{cleveref}
\usepackage{microtype}
\usepackage{gensymb}
\usepackage{float,siunitx}
\usepackage[caption = false]{subfig}

\usepackage[greek,english]{babel}
\usepackage{thumbpdf,enumerate}
\usepackage{booktabs}
\usepackage{sidecap}
\usepackage[scaled=.8]{couriers}
\usepackage{multirow}
\usepackage{placeins}
\usepackage{relsize}
\usepackage{pst-grad,bm}
\usepackage{epigraph}
\usepackage{gensymb}
\usepackage{longtable}
\usepackage{ulem} 
\usepackage{acronym}
\usepackage{physics}
\usepackage{easyReview}

\usepackage[columnwise]{lineno}

\newcommand{\qo}[1]{``#1''}                               		

\DeclareUnicodeCharacter{0301}{\'{e}}

\begin{document}


\vspace{10pt}
\title{Programmable photonic quantum walks on lattices with cyclic, toroidal, and cylindrical topology}

\author{Alessio D'Errico}\email{aderrico@uottawa.ca}
\affiliation{Nexus for Quantum Technologies, University of Ottawa, K1N 5N6, Ottawa, ON, Canada}
\affiliation{National Research Council of Canada, 100 Sussex Drive, Ottawa ON Canada, K1A 0R6}

\author{Nazanin Dehghan}
\affiliation{Nexus for Quantum Technologies, University of Ottawa, K1N 5N6, Ottawa, ON, Canada}
\affiliation{National Research Council of Canada, 100 Sussex Drive, Ottawa ON Canada, K1A 0R6}

\author{Maria Gorizia Ammendola}
\affiliation{Dipartimento di Fisica, Universit\`{a} degli Studi di Napoli Federico II, Complesso Universitario di Monte Sant'Angelo, Via Cintia, 80126 Napoli, Italy}
\affiliation{Scuola Superiore Meridionale, Via Mezzocannone, 4, 80138 Napoli, Italy}

\author{Lukas Scarfe}
\affiliation{Nexus for Quantum Technologies, University of Ottawa, K1N 5N6, Ottawa, ON, Canada}

\author{Roohollah Ghobadi}
\affiliation{Nexus for Quantum Technologies, University of Ottawa, K1N 5N6, Ottawa, ON, Canada}

\author{Francesco Di Colandrea}\email{francesco.dicolandrea@unina.it}
\affiliation{Nexus for Quantum Technologies, University of Ottawa, K1N 5N6, Ottawa, ON, Canada}
\affiliation{Dipartimento di Fisica, Universit\`{a} degli Studi di Napoli Federico II, Complesso Universitario di Monte Sant'Angelo, Via Cintia, 80126 Napoli, Italy}
\author{Filippo Cardano}
\affiliation{Dipartimento di Fisica, Universit\`{a} degli Studi di Napoli Federico II, Complesso Universitario di Monte Sant'Angelo, Via Cintia, 80126 Napoli, Italy}

\author{Ebrahim Karimi}
\affiliation{Nexus for Quantum Technologies, University of Ottawa, K1N 5N6, Ottawa, ON, Canada}
\affiliation{National Research Council of Canada, 100 Sussex Drive, Ottawa ON Canada, K1A 0R6}
\affiliation{Institute for Quantum Studies, Chapman University, Orange, California 92866, USA}

\begin{abstract}
Photonic implementations of unitary processes on lattice structures, such as quantum walks, have been demonstrated across various architectures. However, few platforms offer the combined advantages of scalability, reconfigurability, and the ability to simulate dynamics on lattices with periodic boundary conditions, such as cyclic or toroidal geometries. Here, we employ a recently developed platform that enables the implementation of arbitrary translationally invariant unitary operations on one- and two-dimensional lattices, and demonstrate a natural mechanism for introducing periodic boundary conditions. Our approach leverages direct access to the reciprocal lattice, where discrete sampling of the unitary evolution effectively enforces the desired topology. We program our platform to realize quantum walks on 1D cyclic lattices and 2D lattices with cylindrical or toroidal topologies. The lattice size can be readily tuned by adjusting the sampling density in reciprocal space. By controlling reciprocal-space occupancy, we investigate the dynamics of localized states and wavepackets, observing refocusing behavior, breathing modes modulated by reciprocal-space discretizations, and wavepacket trajectories that reflect the underlying topology. We further demonstrate a form of dimensional reduction by mapping a 2D quantum walk on a cylinder to a 1D walk with a high-dimensional coin. These results establish a versatile platform for realizing a broad class of optical mode transformations within bounded Hilbert spaces.
\end{abstract}

\maketitle 

\section{Introduction}
The spatial and polarization degrees of freedom of light offer an intriguing landscape for classical and quantum information processing. 
The manipulation of a large number of optical modes to perform a variety of tasks is gaining interest in the scientific community as a powerful resource to develop efficient and versatile technologies for artificial intelligence and quantum computing~\cite{lloyd1996universal, aspuru2012photonic, shastri2021photonics}. Two main strategies are being considered: the implementation of integrated optical circuits~\cite{wang2020integrated, bao2023very} and controlled propagation through diffractive optics or complex media~\cite{ehrhardt2021exploring,leedumrongwatthanakun2020programmable,kupianskyi2023high,goel2024inverse,hu2024diffractive, lib2024resource}. In the context of diffractive optics, most implementations have access to a number of modes which is, in principle, unbounded. Nevertheless, some applications could benefit from a system that allows one to choose the dimensions of the modal space and reconfigure the connectivity between the different modes. Some of these operations can be performed with multiplane mode converters~\cite{fontaine2019laguerre, brandt2020high}, where, however, the number of diffractive elements generally depends on the task to be performed and the required accuracy. 

In this work, we consider the implementation of a free-space photonic circuit where the mode space is mapped into a one- or two-dimensional lattice with periodic boundary conditions. Optical modes can thus be spread over a composite lattice with a cycle, cylinder, or torus topology. Optical mode transformations on cyclic graphs have been implemented, for instance, in integrated waveguide systems, and hollow-core fibers~\cite{eriksson2021talbot, hu2025generalized, eriksson2025talbot}. More specifically, coherent transformations involving multiple modes have been extensively considered in quantum walk (QW) applications~\cite{grafe2016integrated,defienne2016two,qiang2024quantum}. A large variety of QW protocols has been introduced over the years. Generally speaking, these can be defined as coherent evolutions of particles on lattices, often, but not exclusively, driven by unitary operations. In the framework of discrete-time QWs, the evolution is driven by translation operators conditioned by the state of some internal degree of freedom (referred to as the \qo{coin}), which cause ballistic propagation phenomena when observed in infinite lattices~\cite{kitagawa2010exploring,schreiber2011decoherence}. QWs have been realized in a wide variety of optical architectures, harnessing all photonic degrees of freedom: optical paths~\cite{broome2010discrete,sansoni2012two,PhysRevLett.112.143604}, time bins~\cite{fenwick2024photonic,monika2025quantum,feis2025space}, frequency~\cite{imany2020probing}, orbital angular momentum~\cite{cardano2015quantum}, and transverse wavevector~\cite{d2020two,DiColandrea2023}. Most of these implementations considered QWs in 1D or 2D infinite lattices, sometimes including boundary effects~\cite{kitagawa2012observation}, as well as spatial and temporal disorder~\cite{schreiber2011decoherence}, and step-dependent operations~\cite{genske2013electric,d2020two,d2021bloch,wang2022high}. 

However, only a few experiments have demonstrated QWs on cyclic lattices, and these have mostly been limited to one-dimensional cycles. Cyclic QWs have been implemented in superconducting qubit cavities~\cite{flurin2017observing}, cascaded beam displacers alternated with waveplates~\cite{bian2017experimental}, optical waveguides~\cite{owens2011two, ehrhardt2021exploring}, time bin implementations with high-dimensional coins~\cite{lorz2019photonic}, and spatial modes~\cite{nejadsattari2019experimental}. To the best of our knowledge, there is a lack of experimental realizations of QWs on higher-dimensional cyclic lattices with toroidal or cylindrical topologies in purely photonic architectures, while some of these topologies have been demonstrated in Rydberg atoms~\cite{barredo2018synthetic} and trapped ion quantum processors~\cite{meth2025simulating}. Here, we extend the capabilities of a recently developed platform to realize QWs on cycles, tori, and cylinders, with arbitrary numbers of lattice sites --limited only by the resolution of the employed devices--. The platform is based on implementing translationally invariant unitary transformations on the polarization-transverse wavevector space. By employing three Spatial Light Modulators (SLMs), which can be controlled by a computer, the platform is fully reconfigurable: the same setup can be used to realize QWs where the number of lattice sites, as well as the number of steps, can be chosen arbitrarily. Unitary protocols can be devised at will and implemented by suitably reconfiguring the phase masks on the SLMs, as demonstrated in a previous work \cite{ammendola2025highdimensionalprogrammablephotoniccircuits}. To implement cyclic QWs, we exploit the fact that our platform gives direct access to the reciprocal lattice space and that periodic boundary conditions can thus be imposed by introducing a discrete sampling of the Brillouin zone. The number of sampled points gives the number of lattice sites, and the far-field intensity distribution can be directly mapped into the particle probability distribution in the corresponding cyclic space. Moreover, by modifying the width of the input spatial mode, we can generate initial states equivalent to wavepackets, and explore how the choice of the central quasi-momentum value affects their dynamics. This work presents proof-of-principle applications, demonstrating refocusing dynamics in 1D and 2D cyclic lattices, wavepacket trajectories, and their interplay with the energy band structure of the underlying protocol.

\section{Theory}
We start by considering the formalism of unitary processes on an infinite lattice, where the action is on a particle equipped with some internal degree of freedom (d.o.f.). The process is described by an operator in the general form \begin{align}
\hat{U}:=\sum_{\textbf{m},\textbf{n}}U_{\textbf{m},\textbf{n}}\otimes\ket{\textbf{m}}\bra{\textbf{n}},
\end{align}
where $U_{\textbf{m},\textbf{n}}$ are matrices acting on the internal d.o.f.. Hereafter, we will assume that each $U_{\textbf{m},\textbf{n}}$ is a $2\times 2$ matrix as our experimental implementation adopts light polarization as the internal d.o.f.. The indices $\{\textbf{m},\textbf{n}\}$ label the lattice sites. Discrete translational invariance implies that $U_{\textbf{m},\textbf{n}}$ can be block-diagonalized in the form
\begin{align}
U_{\textbf{m},\textbf{n}}=\iint_{\text{BZ}} \mathcal{U}(\mathbf{q})e^{i(\textbf{m}-\textbf{n}).\mathbf{q}}\frac{d^Dq}{(2\pi)^D},
\label{eq:Umn}
\end{align}
where $\mathcal{U}(\mathbf{q})$ is a $2\times2$ unitary matrix dependent on the continuous variable $\mathbf{q}$, which is defined within the unit cell of the reciprocal lattice, the Brillouin zone (BZ), and $D$ is the lattice dimensionality --we will consider $D=1,2$. Equation~\eqref{eq:Umn} is equivalent to the following expression for $\hat{U}$:
\begin{align}
\hat{U}=\iint_{\text{BZ}} \mathcal{U}(\mathbf{q})\otimes\ket{\mathbf{q}}\bra{\mathbf{q}}\frac{d^Dq}{(2\pi)^D},
\label{eq:Ucont}
\end{align}
where $\ket{\mathbf{q}}:=\sum_{\textbf{m}}e^{i\textbf{q}\cdot\textbf{m}}\ket{\textbf{m}}$ are the reciprocal lattice eigenstates.
\begin{figure}    \includegraphics[width=\columnwidth]{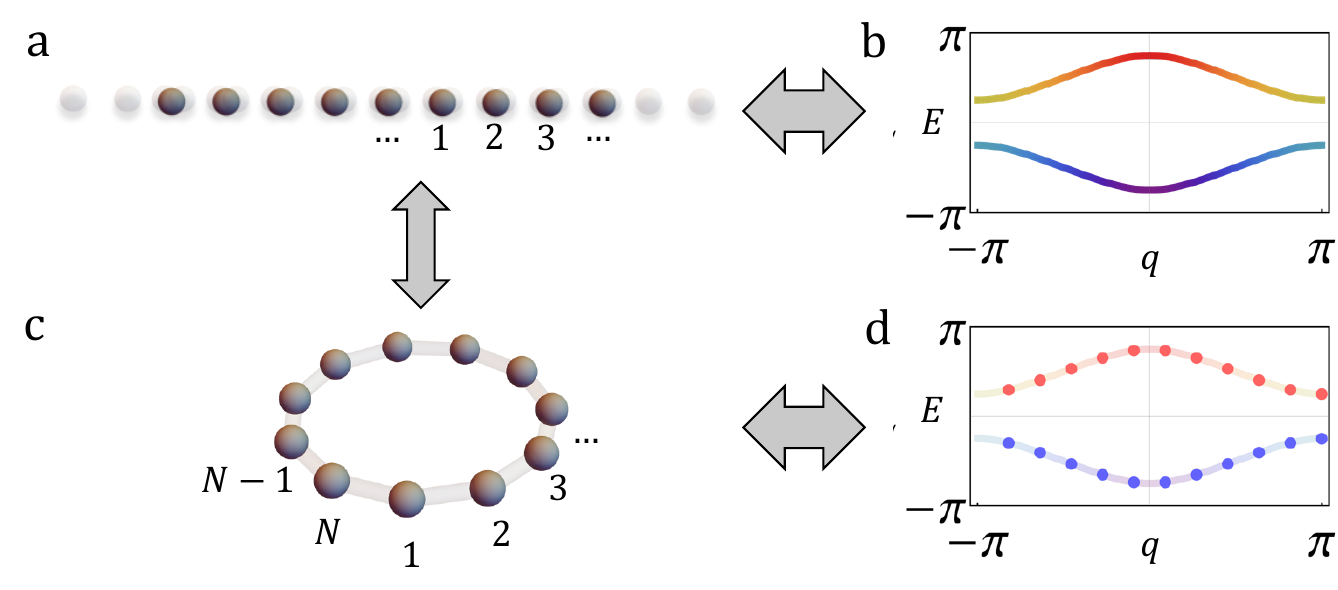}
    \caption{{\bf{Effect of periodic boundary conditions on energy band spectrum.}} An infinite lattice with translationally invariant interatomic interactions (a) has energy eigenvalues $E(q)$ labeled by the continuous variable $q$, which is defined in a finite interval, the Brillouin zone (b). If, instead of an infinite lattice, one considers $N$ sites with periodic boundary conditions (c), then the eigen-energy depends on a discrete subset of $N$ points in the Brillouin zone (d). The energy band structure plotted here corresponds to the $\hat{U}_{1D}$ protocol considered in this work.}
    \label{fig:spectrum}
\end{figure}

Thus far, we assumed an infinite lattice. Cyclic lattices, with $N$ sites along a specified direction, can be modeled by imposing periodic boundary conditions, i.e., the evolution operator and the corresponding eigenstates are mapped into themselves by a translation of $N_{\mathbf{t}}$ lattice sites along the translation vector ${\mathbf{t}}$. Thus, for each ${\mathbf{t}}$, we impose the following constraint on the plane wave contribution in the Bloch wavefunctions: $\braket{m_{\mathbf{t}}}{q_{\mathbf{t}}}=\braket{m_{\mathbf{t}}+N_{\mathbf{t}}}{q_{\mathbf{t}}}$, which implies $1=\exp(i N_{\mathbf{t}} q_{\mathbf{t}})$, hence $q_{\mathbf{t}}=2\pi h/N_{\mathbf{t}}$, where $h=1,\ldots, N_{\mathbf{t}}$. In this way, the quasi-momentum becomes a discrete variable and the BZ contains the same number of reciprocal lattice sites as the direct lattice. The corresponding unitary then reads
\begin{align}
\hat{U}:=\frac{1}{N_T}\sum_{\textbf{h}}\mathcal{U}(q_{\textbf{h}})\otimes\ket{q_{\textbf{h}}}\bra{q_{\textbf{h}}},
\label{eq:Ucyclic}
\end{align}
where $\mathbf{h}=1,\ldots, N$ for $D=1$, and $\mathbf{h}=(h_x,h_y)$ with $h_{x,y}=1,\ldots,N_{x,y}$ for $D=2$ (in the case of square lattices), and $N_T$ is the number of lattice sites. 

\begin{figure*}    \includegraphics[width=\textwidth]{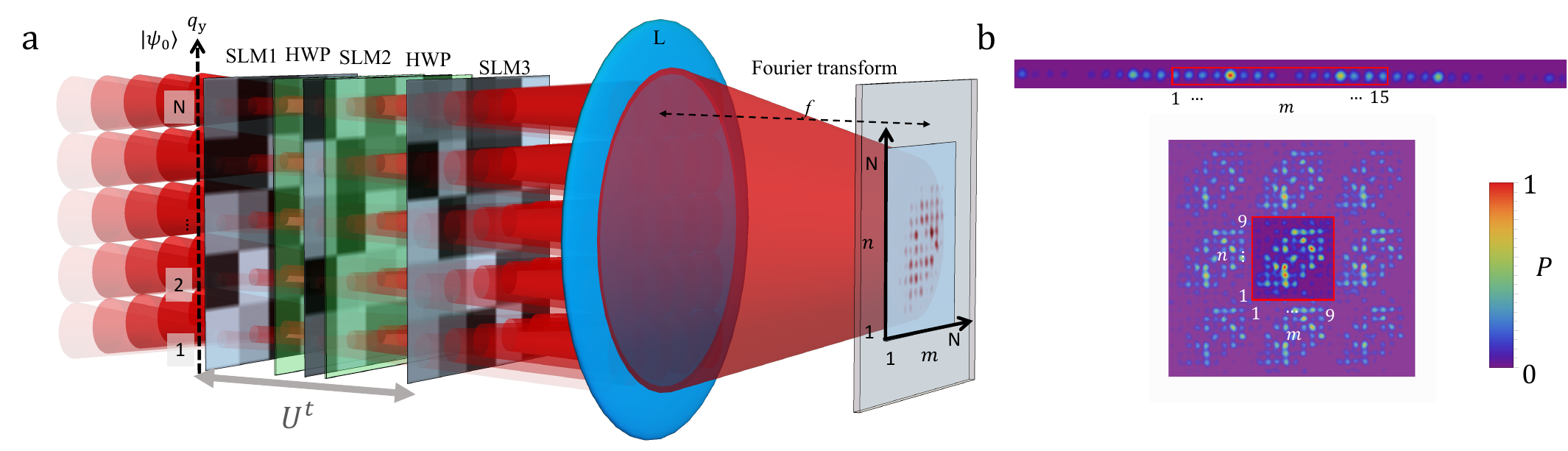}
    \caption{\textbf{Experimental scheme.} (a) Concept of the experiment. An input laser beam with transverse intensity consisting of a regular array of narrow spots (beamlets) is incident on a stack of spatially dependent polarization-transforming devices which effectively implement a unitary transformation $U^t$ in the photonic spin-momentum space. The stack is realized with three liquid-crystal spatial light modulators (SLMs), each displaying discrete phase masks with the same arrangement of the input light spots. The axis of the intermediate SLM is effectively rotated by $45^{\circ}$ by means of two half-wave plates (HWP). The resulting transformation corresponds to Eq.~\eqref{eq:Ucyclic}, with $\mathbf{q}_{\mathbf{h}}$ mapped into transverse spatial coordinates $(x_{h_x},y_{h_y})$. (b) Examples of intensity patterns collected in the far field for quantum walks on a cycle with 15 lattice sites (top inset) and a torus with 9$\times$9 lattice sites (bottom inset). Note that the patterns exhibit multiple replicas of the QW process. The relative intensities of these replicas is modulated by an envelope function that depends on the width of the input beamlets.}
    \label{fig:simplified setup}
\end{figure*}

\begin{figure*}    \includegraphics[width=\textwidth]{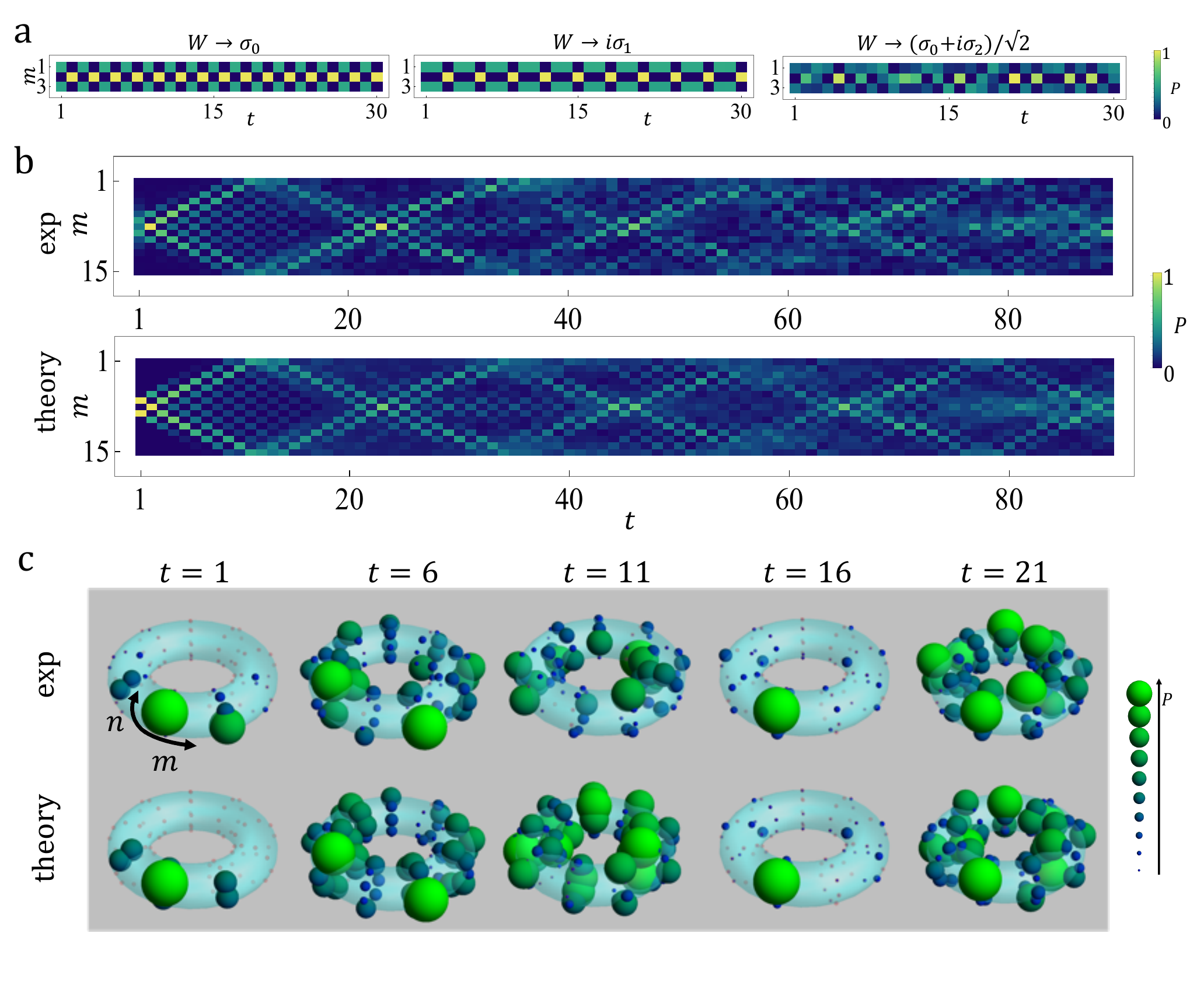}
    \caption{\textbf{QWs of initially localized states on cyclic 1D and 2D lattices.} (a) Experimental recurrent dynamics on a lattice with 3 sites. We choose different modifications of the $\hat{U}_{1D}$ protocol. If the coin operator $W$ is replaced with a Pauli matrix (here, the cases with $\sigma_0$ and $\sigma_1$ are shown), then the QW exhibits perfectly recurrent dynamics with the period given by, respectively, 2 and 3 steps. The original coin operator $W=(\sigma_0+i\,\sigma_2)/\sqrt{2}$ yields more complex dynamics instead (almost perfect recurrence is expected after 253 steps). (b) Experimental and theoretical results of $\hat{U}_{1D}$ applied for up to 90 steps on an initial state prepared as $\ket{m=7}\otimes\ket{H}$. (c) Experimental and theoretical results of $\hat{U}_{2D}$ applied for up to 21 steps on an initial state prepared as $\ket{m=n=0}\otimes\ket{H}$. Extended experimental results are shown in the Methods. Similarity values are: $S=0.9998\pm0.0002$ for $W\rightarrow\sigma_0$ and $W\rightarrow i\sigma_1$, $S=0.97\pm0.03$ for $W\rightarrow (\sigma_0+i\,\sigma_1)/\sqrt{2}$ in panel (a), $S=0.92\pm0.05$ in (b), and $S=0.94\pm0.02$ in (c), where the errors correspond to the standard deviation of the similarity values.}
    \label{fig:locstates}
\end{figure*}

\begin{figure*}
\includegraphics[width=\textwidth]{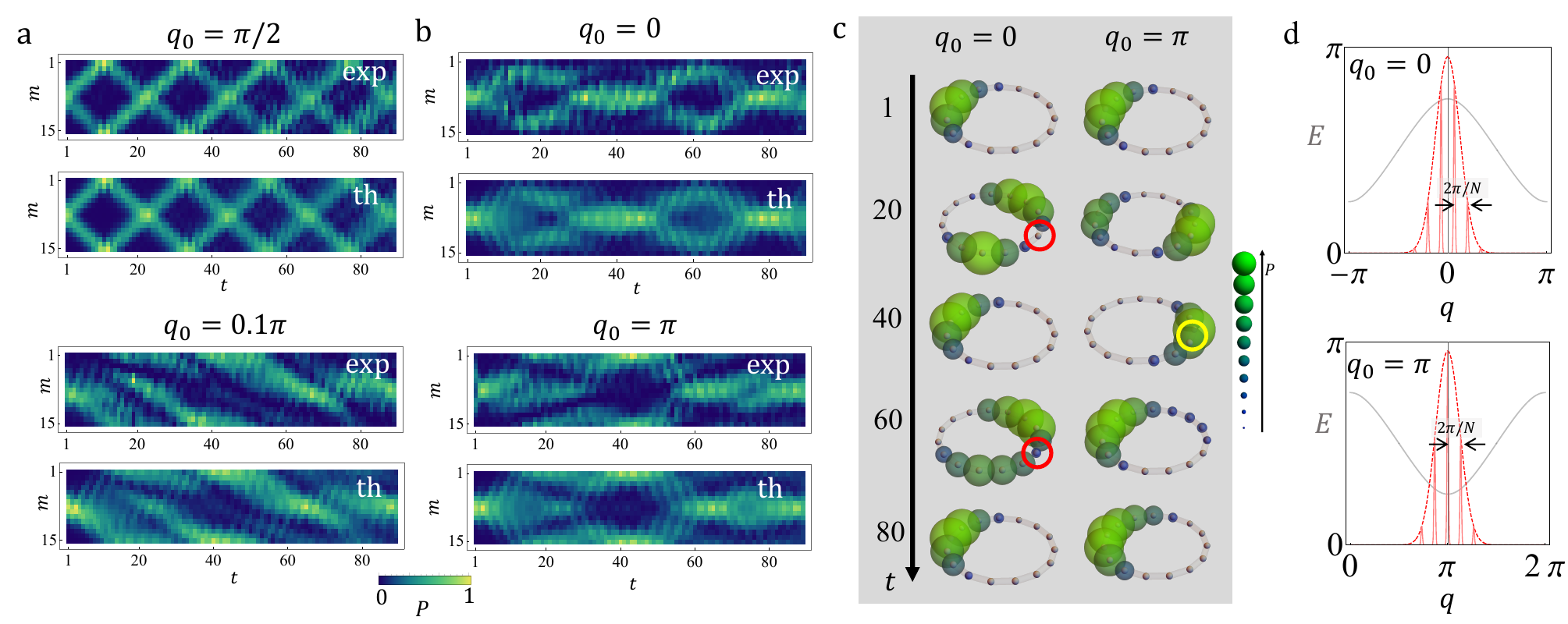}
    \caption{\textbf{QWs of wavepackets on a 1D cycle}. (a) Evolution of wavepackets with non-zero group velocity on a lattice with 15 sites. (b) Evolution of wavepackets with zero group velocity on a lattice with 15 sites. Different breathing phenomena are observed depending on whether $q_0$ is an allowed quasi-momentum eigenstate or not. (c) These different features are highlighted by showing snapshots of the probability distributions every $20$ time steps. The antipodal point is marked by red or yellow circles at those time-steps where the localization at that point is avoided or occurs. (d) The initial states considered in (b) and (c) can be decomposed into an even or odd number of quasi-momentum eigenstates for $q_0=0$ and $\pi$, respectively. The superposition of these states, which are plane waves with wavevectors $q_h$, leads to interference effects responsible for the observed breathing phenomena.  Similarity values are: $S=0.98\pm 0.01$ for $q_0=\pi/2$, $S=0.98\pm 0.02$ for $q_0=0.1\pi$, $S=0.98\pm 0.01$ for $q_0=0$, and $S=0.97\pm 0.02$ for $q_0=\pi$. Errors correspond to the standard deviation of the similarity values.}
    \label{fig:1dwavepackets2}
\end{figure*}

\textit{Energy discretizations and refocusing dynamics.} In Fig.~\ref{fig:spectrum}, we illustrate the effect of imposing periodic boundary conditions on the eigenvalue spectrum of $\hat{U}$. In the infinite-lattice case, the spectrum is readily found from the SU$(2)$ representation of $\hat{U}$: $\mathcal{U}(\mathbf{q})=\exp(i E(\mathbf{q}) \mathbf{n}(\mathbf{q})\cdot\boldsymbol{\sigma})$, where $\boldsymbol{\sigma}=(\sigma_1,\sigma_2,\sigma_2)$ is a vector formed by the three Pauli matrices, $\mathbf{n}(\mathbf{q})$ corresponds to a unit vector coincident with the Bloch sphere representation of the eigenstates of $\mathcal{U}(\mathbf{q})$, and $E(\mathbf{q})$ is the phase of the eigenvalues of $\mathcal{U}(\mathbf{q})$, which is often denoted as \textit{quasi-energy}. Being continuous functions of the quasi-momentum, the quasi-energies $\pm E(\mathbf{q})$ form a band structure (see Fig.~\ref{fig:spectrum}-a,b), and the dimension of the internal d.o.f. determines the number of bands. From Eq.~\eqref{eq:Ucyclic}, one immediately sees that the effect of the periodic boundary conditions is to limit the set of available eigen-energies to $E(\mathbf{q}_{\mathbf{h}})$ (see Fig.~\ref{fig:spectrum}-c,d), with important consequences on the single-particle dynamics.

In this work, we will focus on implementing QW processes on lattices where periodic boundary conditions are imposed in at least one spatial direction. In QWs, the same unitary process is applied multiple times, defined as (discrete) time steps: $\hat{U}_{QW}(t):=\hat{U}^t$, with $t\in \mathbb{N}$. In this case, each eigenstate takes a phase $\exp(i E(\mathbf{q_h})t)$ after $t$ steps, and the evolved state can generally be written as 
\begin{equation}
\ket{\psi(t)}=\sum_{\mathbf{h},\sigma=\pm}C_{\mathbf{h},\sigma}e^{i \sigma E(\mathbf{q_h})t}\ket{\mathbf{n}_{\sigma}(\mathbf{q}_{\mathbf{h}})}\otimes\ket{\mathbf{q}_{\mathbf{h}}},
\label{eq:QWstate}
\end{equation}
where the complex coefficients $C_{\mathbf{h},\sigma}$ specify the initial state.
The representation of the lattice space can be obtained as $\psi(\mathbf{m},t):=\braket{\mathbf{m}}{\psi(t)}$, using $\braket{\mathbf{m}}{\mathbf{q}_{\mathbf{h}}}=\exp(i\mathbf{m}\cdot \mathbf{q}_{\mathbf{h}})$:
\begin{equation}
\psi(\mathbf{m},t)=\sum_{\mathbf{h},\sigma=\pm}C_{\mathbf{h},\sigma}e^{i (\sigma E(\mathbf{q_h})t+\mathbf{m}\cdot \mathbf{q}_{\mathbf{h}})}\ket{\mathbf{n}_{\sigma}(\mathbf{q}_{\mathbf{h}})}.\label{eq:QWstatelattice}
\end{equation}
Note that from this expression, it is straightforward to verify that the periodic boundary conditions are satisfied: $\psi(\mathbf{m}+\mathbf{N})=\psi(\mathbf{m})$.

The finite number of available eigen-energies allows for the possibility that there exists a time step $t^*$ such that $E(\mathbf{q}_{\mathbf{h}})t^*=2n_{\mathbf{h}}\pi$, with integer $n_{\mathbf{h}}$, which is equivalent to $\hat{U}^{t^*}=\mathbb{I}$~\cite{dukes2014quantum}. However, it suffices for two of the system eigenvalues to be incommensurate ($E(\mathbf{q}_{\mathbf{h}_1})/E(\mathbf{q}_{\mathbf{h}_2})\neq n_1/n_2$) for this condition to not be verified. Nevertheless, it can easily happen that the system returns very close to the initial state due to the compactness of the space, as we will show in some examples. Hence, while in QWs on infinite lattices the wavefunction typically exhibits ballistic propagation, in cyclic lattices, refocusing phenomena are widespread.

Another consequence of the discretizations of the energy spectrum is that the group velocity, given by $\mathbf{v}_g(\mathbf{q})=\nabla_{\mathbf{q}}E(\mathbf{q})$ in the continuous case, cannot be rigorously defined. However, as we will show experimentally, it still proves to be a useful concept in gaining a heuristic understanding of the dynamics of wavepackets, also for relatively small values of $N$. Interestingly, new behaviors can appear when considering wavepackets prepared with initial momentum corresponding to maxima or minima of the energy spectrum. As we show in the following section, refocusing dynamics can be observed with a periodicity that depends on whether the extrema of the continuous spectrum are sampled or not when moving to the discrete case. We show that this effect depends on the interplay between the wavepacket spreading and the limited number of lattice sites, which leads to self-interference phenomena.

\section{Experiment}

The experimental platform, outlined in Fig.~\ref{fig:simplified setup}-a (and detailed in Methods A and Supplementary Figure S2), consists of three cascaded liquid-crystal-based SLMs, arranged to allow for the implementation of spatially dependent polarization transformations. In this platform, the transverse coordinates in the SLMs plane correspond to the position in the reciprocal lattice, which gives us the possibility of directly implementing transformations in the form of Eq.~\eqref{eq:Ucont}. Indeed, in a previous work~\cite{ammendola2025highdimensionalprogrammablephotoniccircuits} it was demonstrated that this architecture can be used to implement arbitrary unitaries such as those of Eq.~\eqref{eq:Ucont}, in analogy with the approach shown in Ref.~\cite{DiColandrea2023} in the case of tailored Phancharatnam-Berry optical elements. Each SLM can be described as a wave retarder with fixed extraordinary axis and variable optical retardation $\Delta(x,y)$. For instance, an SLM with an extraordinary axis parallel to horizontal polarization can be modeled as $\hat{SLM}=e^{i\Delta(x,y)}\ket{H}\bra{H}+\ket{V}\bra{V}$, where $\ket{H}$ and $\ket{V}$ correspond to horizontal and vertical polarization states, respectively. Eq.~\eqref{eq:Ucont} is implemented via the transformation 
$\hat{U}=\hat{SLM_3}\cdot \hat{HWP}(-22.5 \degree)\cdot \hat{SLM_2}\cdot \hat{HWP}(22.5\degree)\cdot \hat{SLM_1}$, where $\hat{HWP}(\alpha)$ is a half-wave plate with an extraordinary axis at an angle $\alpha$ with respect to the horizontal axis,
and each SLM operator is characterized by a different retardation pattern $\Delta_i(x,y)$ $(i=1,2,3)$. To realize periodic boundary conditions, we can straightforwardly calculate the patterns $\Delta_i(x,y)$ (see Methods B) that implement the unitary $\hat{U}$ only for those values of $(x,y)$ corresponding to the sampled positions $\mathbf{q}_{\mathbf{h}}$ in the reciprocal space. Ideally, light should be incident only on the pixels corresponding to the allowed quasi-momentum values. The incident beam on the SLM stack should thus appear as a regular grid of narrow light spots (beamlets), spaced with a period given by $\Lambda/N$, with $\Lambda$ corresponding to the BZ size calculated on the SLM plane. Each beamlet should have a narrow diameter. Details on how the patterns were implemented and the beamlets prepared are given in the Method B and Supplementary Figure S2.  

The lattice space is directly accessed via a 2D optical Fourier transform, i.e., by measuring the light field in the focal plane of a lens placed at the end of the SLMs stack. Examples of retrieved far-field intensity distributions are shown in Fig.~\ref{fig:simplified setup}-b. One typically observes, in the far field, separate multiple \textit{replicas} of the expected distribution on the cyclic lattice. The number of copies should be, in principle, infinite if the input beamlets are infinitely narrow. In practice, this \qo{infinite} pattern is multiplied by an envelope given by the Fourier transform of the beamlet transverse amplitude (see Methods B). The beamlets' waist should then be chosen in such a way that their far-field representation is sufficiently flat in the region corresponding to the expected number of lattice sites in the cyclic QW.  

In our experiment, we implement the cyclic version of the 1D and 2D QW protocols already considered in Refs~\cite{cardano2017detection, d2020two}. These protocols are based on two simple building blocks: 1) a polarization-dependent translation on the lattice: $T_{\xi}:=\cos(\delta/2)\mathbb{I}+i\sin({\delta/2})\sum_\xi{(\ket{\xi+1}\bra{\xi}\otimes\sigma_-+h.c.)}$, where $\xi$ indicates any direction along the translation vectors of the lattice, $\sigma_-=\frac{1}{2}(\sigma_1 - i \sigma_2$), and $\delta$ is a tunable parameter; 2) a polarization rotation, often called \textit{coin operator}, e.g., $W=(\sigma_0+i\sigma_2)/\sqrt{2}$. For 1D cyclic processes, we mainly consider the protocol $\hat{U}_{1D}=\hat{W}\cdot\hat{T}_x(\pi)$, while the process considered on a 2D square lattice is $\hat{U}_{2D}=\hat{T}_y(\pi/2)\cdot\hat{T}_x(\pi/2)\cdot\hat{W}$. These protocols have been studied in previous QW implementations, particularly in connection with their topological properties~\cite{cardano2015quantum,cardano2017detection,DErrico2020,d2020bulk,di2024manifestation,ammendola2025large}. The corresponding energy spectra are given in Methods C. However, we point out that we can choose in our platform arbitrary unitaries with discrete translational symmetry~\cite{DiColandrea2023, ammendola2025highdimensionalprogrammablephotoniccircuits}. 

\subsection{Evolution of initially localized states}
Figure~\ref{fig:locstates} shows the experimental results of the dynamics of states initially localized in the lattice space. As first examples, we investigated the case of ${D=1}$ and ${N=3}$ (see Fig.~\ref{fig:locstates}-a). In this scenario, there are three allowed quasi-momentum eigenvalues, $q_h=2\pi h/3$, each with two associated quasi-energies $\pm E(q_h)$, with values depending on the chosen protocol. We show the dynamics for three protocols where the energy eigenvalues are identical, commensurate, or quasi-incommensurate. The first two cases show a perfectly recurrent process with period $t^*=2,3$, respectively, while the last presents a more complex dynamics having a characteristic period ${t^*\sim253}$ (details on the protocols are reported in the Methods D).

To explore dynamics on larger values of $N$, we realized 90 steps for the $\hat{U}_{1D}$ protocol on a lattice with 15 sites (see Fig.~\ref{fig:locstates}-b) and 30 steps for the 2D protocol on a square lattice with 9$\times$9 lattice sites (see Fig.~\ref{fig:locstates}-c). Note that periodic boundary conditions on a 2D square lattice endow it with a torus topology. The first steps are equivalent to infinite lattice dynamics, showing the characteristic ballistic spreading. At larger step numbers, trajectories evolving in opposite directions merge and interfere, thus altering the probability distributions at successive steps. As a consequence, there can be instances where the probability distribution is strongly peaked at the starting position. In the 1D scenario, we observe partial refocusing of the probability distribution after $t=23$ steps, whereas in the 2D case, this occurs at $t=16$. The refocusing is partial since none of the time steps considered in these examples corresponds to an identity operator. Throughout the paper, we quantify the agreement with the theory via the similarity,  $S=(\sqrt{\mathbf{P}_\text{exp}}\cdot \sqrt{\mathbf{P}_\text{th}})^2$, where $\mathbf{P}_\text{exp,th}$ are unit vectors obtained by flattening the normalized experimental and theoretical distributions, respectively. We observe similarities of the order of 90$\%$. Deviations from theory can be ascribed mainly to finite size of the incident beamlets and imperfections in the preparations of the initial state (as described in Method B).  
\begin{figure*}
\includegraphics[width=\textwidth]{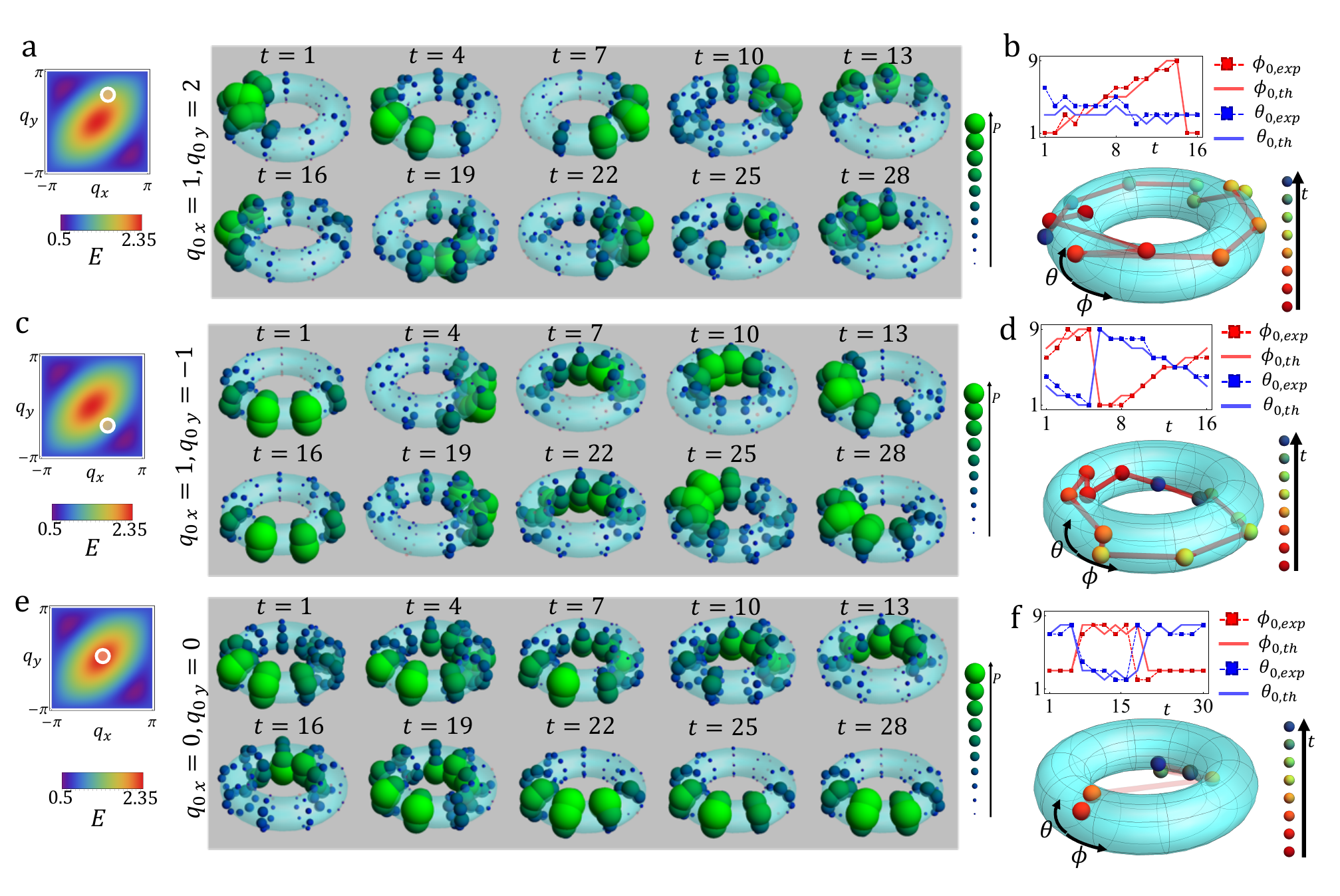}
    \caption{\textbf{QWs of wavepackets on a 2D cycle.} Snapshots of the probability distributions of wavepackets on the torus are shown in (a), (c), and (e), for different values of $\mathbf{q}_0$, illustrated in the insets plotting the upper energy band. (a) A wavepacket motion occurring mainly on the $\phi$ direction is considered. (b) Trajectory of the maximum of the probability distribution as 1D plots of the $\theta_0$ and $\phi_0$ coordinates (top inset), compared with the theoretical ones (continuous curves), and visualized directly on the torus (bottom inset). (c)-(d).~Results for a wavepacket tracing a trajectory that wraps around the torus in 16 steps. (e)-(f).~Wavepacket motion with zero group velocity, where the wavepacket tunnels periodically between opposite regions of the lattice. Similarity values are: $S=0.88\pm0.8$, $S=0.94\pm0.03$, and $S=0.91\pm0.05$, for the results in (a), (c), and (e), respectively. A detailed comparison between experimental and theoretical results is provided in the Supplementary Figure S3.}
    \label{fig:2dwavepackets2}
\end{figure*}
\begin{figure*}
\includegraphics[width=\textwidth]{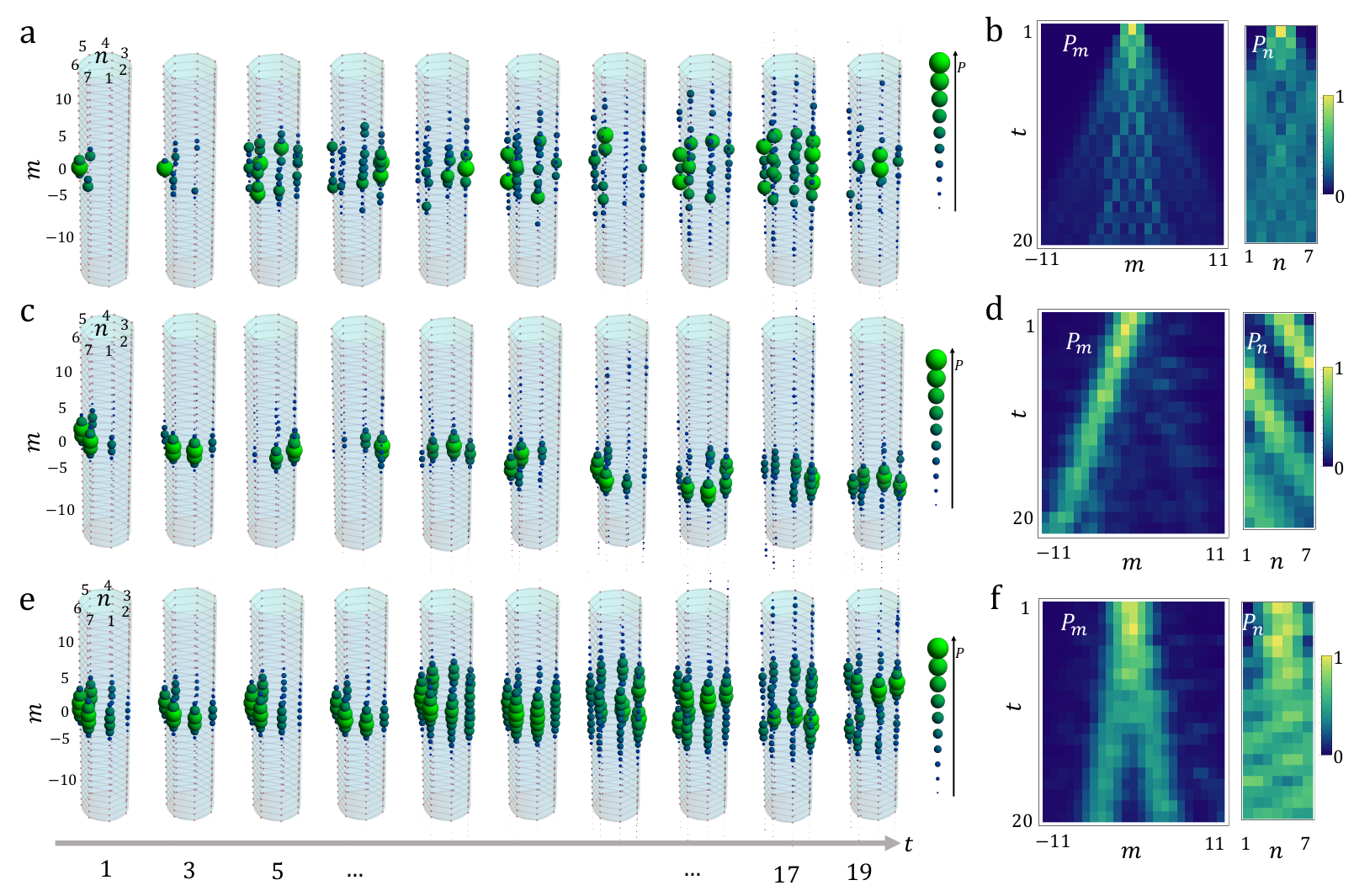}
    \caption{\textbf{QWs on a cylinder.} Snapshots of the probability distributions on the cylinder are shown for a localized state (a) and wavepackets (c)-(e). (b) Marginal distributions, $P_m$ and $P_n$, along the infinite (left) and cyclic (right) coordinates, with $P_m$ showing a characteristic ballistic propagation with several peaks, reminiscent of QWs with a high-dimensional coin d.o.f.. These multiple peaks can be attributed to fast and slow wavepacket contributions, as the ones considered in (c)-(d) and (e)-(f), respectively. (c)-(d)~Dynamics of a wavepacket initialized in $(q_x,q_y)=(-1.57,-1)$, which evolves wrapping around the cylinder, as evident from the marginal probabilities in (d). (e)-(f)~Dynamics of a wavepacket initialized in $(q_x,q_y)=(\pi,\pi)$, where the marginal distribution $P_m$ shows a ballistic spreading slower than in d. Similarity values are:  $S=0.93\pm0.04$, $S=0.94\pm0.04$, and $S=0.955\pm0.02$.
    A detailed comparison between experimental and theoretical results is provided in the Supplementary Figure S4.}
    \label{fig:cylinder}
\end{figure*}

\subsection{Wavepacket dynamics}

In this section, we consider initial states occupying more than one lattice site and having uniform polarization. If the wavepacket's width in the lattice space is sufficiently large, these states occupy a narrow region in the reciprocal space, and their dynamics is thus affected by the quasi-energy values in that region. The experimental results of the wavepackets were prepared by reducing the aperture of a pinhole placed before the first SLM. Adjusting the center position of the pinhole in the transverse plane, we could prepare wavepackets with different initial quasi-momentum $q_0$. The results are in agreement with the model of Gaussian wavepackets with a width $w$ (defined below) of $2-2.8$ lattice sites.

\textit{Wavepackets on 1D cycles.}
In an infinite lattice, the evolution of a Gaussian wavepacket in the form $\ket{\psi_0(q_0,w)}\propto\ket{s}\otimes\sum_m \exp(-(m-m_0)^2/w^2-iq_0 m)\ket{m}$, of width $w$, polarization $\ket{s}$, and average quasi-momentum $q_0$, can be described in terms of the group velocity and group velocity dispersion. The state generally splits into two wavepackets, each with relative amplitude $\abs{\braket{s}{\pm\mathbf{n}(q_0)}}$, propagating with speed $\pm v_g(q_0)$, and the width spreads proportionally to the energy band curvature. If the two wavepackets are sufficiently narrow, their polarization state keeps being approximately $\ket{\pm\mathbf{n}(q_0)}$ during the propagation. In a cyclic lattice, the two portions of the wavepackets will return to the original position and cross each other essentially undisturbed, having approximately orthogonal polarization. This is observed in Fig.~\ref{fig:1dwavepackets2}-a for the case $q_0=\pi/2$ corresponding to maximum group velocity and minimal width spreading in the spectrum of the protocol $\hat{U}_{1D}$. The two wavepackets travel through the cyclic lattice with opposite group velocities. At larger step numbers, it is possible to notice the weak spreading of the wavepackets' width due to the group velocity dispersion. This effect is more pronounced for $q_0$ closer to the extrema of the energy spectrum (e.g., $q_0=0.1\pi$ shown in the bottom inset of Fig.~\ref{fig:1dwavepackets2}-a). A perhaps more surprising phenomenon is observed when considering wavepackets prepared with $q_0$ corresponding to $v_g=0$ (see Fig.~\ref{fig:1dwavepackets2}-b,c). This corresponds to values of $q_0$ where the energy gap exhibits a maximum or minimum. In these cases, the wavepacket is expected to stay centered at the origin and to spread in width due to the band curvature being maximum at these points. For a cyclic lattice, the state can spread enough to occupy all the lattice sites and consequently will start to interfere with itself. This self-interference effect can induce the wavepacket to become suddenly localized in a position antipodal to the initial one. This is observed for $q_0=\pi$ where, after $\sim 40$ steps, the wavepacket, initially prepared around $m_0=7$, appears peaked in the opposite lattice position $m_0=1$, and eventually refocuses back in $m_0=7$ after $\sim 40$ additional steps. A different breathing phenomenon happens when $q_0=0$.  In this case, at some time steps, the wavepacket never occupies the antipodal site around $m=1$, while the occupancy of the initial central position oscillates in time. Moreover, the refocusing happens with a period halved with respect to the previous case. The different behavior at $q_0=0$ and $\pi$ is at first puzzling when trying to interpret the results from the continuous energy spectrum. Indeed, in the case of the specific protocol considered, these two points are perfectly equivalent in the infinite lattice case (this is a consequence of the quasi-energy being defined modulo $2\pi$). The difference arises from the discrete sampling of the BZ and on whether $q_0$ is an allowed momentum eigenvalue or not. By closely inspecting the spectrum in Fig.~\ref{fig:spectrum}-(d) or the plots in Fig.~\ref{fig:1dwavepackets2}-(d), we realize that, while $q_0=\pi$ is an allowed momentum eigenvalue, $q_0=0$ is not. We can decompose the initial state into the eigenstates of the system, and truncate the decomposition at some desired order. For $q_0=\pi$, an odd number of terms is needed in the expansion, while for $q_0=0$, an even number is needed, as illustrated in Fig.~\ref{fig:1dwavepackets2}-(d). Moreover, if the wavepacket in the reciprocal space is narrow enough, we can assume that the eigenstates of the decomposition have the same polarization. As shown in Method D, comparing the interference of, respectively, only 4 and 3 waves, is sufficient to capture the observed phenomenology. The destructive/constructive interference effects can be predicted by comparing the energy differences between the first few eigenmodes contributing to the wavepacket expansion. For $q_0=0$, one should consider the difference $\delta E_{0}=E(\pi/N)-E(3\pi/N)$, while, for $q_0=\pi$, one should consider the difference $\delta E_{\pi}=E(\pi)-E(\pi-2\pi/N)$. In our protocol, $\delta E_0\approx 2\delta E_{\pi}$ which explains why we observe a doubled refocusing time for $q_0=\pi$ with respect to $q_0=0$. For $q_0=0,\pi$ the refocusing time is expected to be $t_{0,\pi}=2\pi/\delta E_{0,\pi}$, which gives $(t_0,t_{\pi})=(40,76)$. The behavior at the antipodal site ($m=1,N$) can also be explained in terms of the interference of 3 or 4 plane waves. For $q_0=0$, the wavepacket is approximated by $\psi_0(m,t)=c_1\cos\big((m-m_0)\pi/N \big)+c_3\cos\big((m-m_0)3\pi/N \big)\exp(i\delta E_0 t)$, where the coefficients $c_{1,3}$ are given by the Gaussian envelope and the arguments of the cosines are chosen in such a way that the initial state is peaked around $m_0\sim N/2$. Thus, $\psi_0(N,t)\approx0$, since $|m-m_0|\approx N/2$ for $m=N$, which explains why, for this case, we never observe a localization in the antipodal lattice site. Similar arguments show that localization in the antipodal site, $m=1,N$, happens periodically for $q_0=\pi$ (detailed calculations are outlined in Method D). Adding more terms in the expansion of $\psi_{0,\pi}$ allows for the explanation of finer details of the wavepacket propagation (see Methods D and Supplementary Fig. S1), except for small interference fringes which require including the smooth quasi-momentum variation of the polarization-eigenstates. This breathing phenomenon could be observed in other platforms, such as Modulo-OAM sorters based on hollow-core fibers \cite{hu2025generalized} and 3D integrated waveguides.

\textit{Wavepackets on the torus.} We now consider the QW of a wavepacket on a discrete torus with 9$\times$9 lattice sites, labeled by angular indexes $(\theta, \phi):=\frac{2\pi}{N}(m,n)$. In a torus, wavepackets with non-zero group velocity can follow topologically distinct trajectories. For instance, Figure~\ref{fig:2dwavepackets2}-(a),(b) shows the experimentally measured dynamics of a wavepacket with mean group velocity pointing mainly along the $\phi$ direction (apart from a small component of $\mathbf{v}_g$ along $\theta$ due to experimental imperfections). In the first 16 steps, the wavepacket's center of mass traces a trajectory close to the torus longitude curve. In Figs.~\ref{fig:2dwavepackets2}-(c),(d) we show results where the group velocity has equal components along $\theta$ and $\phi$ and the wavepacket's center of mass follows a trajectory that winds around the torus with a period of 16 time steps, which is topologically equivalent to a Villarceau circle (a geodesic obtained by obliquely intersecting a torus with a plane). Both of these cases were realized with a wavepacket prepared with horizontal polarization and a width of $\sim 2.3$ lattice sites.
We also observe the dynamics in the case of a wavepacket with zero group velocity (see Figs.~\ref{fig:2dwavepackets2}-(e),(f)). Similarly to the 1D case, the group velocity dispersion and the boundedness of the lattice space compete so as to induce the tunneling of the wavepacket's center of mass from the initial position to the opposite one, and back, as shown in Fig.~\ref{fig:2dwavepackets2}-(f).

\subsection{QW on a cylinder}

In the case of a process with $D>1$, periodic boundary conditions can be imposed only along one direction. Here we consider the case where the $x$ axis (with lattice sites labeled by the index $m$) is unbounded, $m\in\mathbb{Z}$, while the $y$ axis is a cyclic coordinate ($n={1,\ldots, N_y}$). The lattice space is thus topologically equivalent to a cylinder. This kind of process can also be mapped in a 1D quantum walk along the $x$ direction with a high-dimensional coin space, where the dimensionality is $d_c=2N_y$ -- with the factor of 2 due to the polarization d.o.f.--~\cite{bru2016quantum}. The effective 1D dynamics is thus regulated by a quasi-energy spectrum given by $d_c$ continuous quasi-energy bands, and, as a consequence, multiple group velocity values are possible for a fixed quasi-momentum $q_x$. The probability distribution of the effective 1D model, $P_m$, can be extracted as a marginal of the full probability density $P=P_{m,n}$ on the cylinder: $P_m:=\sum_{n=1}^{N_y}P_{m,n}$. As in the case of two-level coins, these processes still exhibit ballistic evolution; however, the probability distributions can display multiple peaks due to the several values of allowed group velocities~\cite{lorz2019photonic}. This is qualitatively evident in Figs.~\ref{fig:cylinder}-(a),(b), which show the evolution of a localized state. Panel (b) shows the marginal probability distributions ($P_n=\sum_m P_{m,n}$), with $P_m$ showcasing a ballistic spreading with $\approx 2$ characteristic velocities. This effect can be attributed to different wavepackets participating in the initial state decomposition. As demonstrated in Ref.~\cite{lorz2019photonic}, the peaks are associated with wavepackets featuring extrema of $v_g$. The corresponding dynamics, shown in Figs.~\ref{fig:cylinder}-(c),(d), display a probability distribution whose maximum propagates wrapping around the cylinder. The marginal probabilities in Fig.~\ref{fig:cylinder}-(d), $P_m$ and $P_n$, highlight the mean wavepacket's velocity along the open and cyclic directions, respectively. Lower propagation speeds are induced by other contributions, such as the wavepacket prepared in $(q_x,q_y)=(\pi,\pi)$, which evolves propagating mainly along the $x$ direction, as evident from the marginals in Fig.~\ref{fig:cylinder}-f. Note that, in this example, the wavepacket splits into two due to the initial polarization ($\ket{H}$) being approximately an equal superposition of the local polarization eigenstates. 
This example shows how our capability to induce periodic boundary conditions in our photonic mode mixer allows for the possibility of exploring ideas connected with the concept of dimensional reduction, such as the classification of high-order topological insulators~\cite{PhysRevB.78.195424}, bulk measurements of topological invariants in multi-band systems~\cite{maffei2018topological, d2020bulk}, and physical effects of non-Abelian gauge theories~\cite{klco20202}.

\section{Conclusions}

We demonstrated a simple strategy to implement periodic boundary conditions in simulators of lattice dynamics based on photonic synthetic dimensions. If the simulator relies on direct access to the reciprocal lattice space, one can impose the desired boundary conditions by adequately exciting a regular subset of points in the Brillouin zone. This approach could be extended to other platforms, such as those employing Bose-Einstein condensates~\cite{dadras2018quantum}, time bins, and frequency domains. Our platform, consisting of three reconfigurable pixelated elements performing transformations in the polarization and transverse wavevector space of light~\cite{he2025reconfigurable,ammendola2025highdimensionalprogrammablephotoniccircuits}, can be used to implement QWs on cycles, tori, and cylinders, with the possibility of programming the underlying unitary process, the number of time steps, and the state initialization. We demonstrated how the cyclic structure of the lattice leads to recurrent phenomena, as well as breathing effects deriving from the interplay between group velocity dispersion and the sampling of the reciprocal space. Some of these effects can also be observed in hollow-core fibers. We similarly engineered wavepacket dynamics on toroidal and cylindrical lattice spaces with a rich variety of trajectories. Finally, we interpreted the QW on a cylinder from the point of view of dimensional reduction, showing the equivalence with a 1D QW with a high-dimensional coin degree of freedom. This allowed us to extend structured light-based architecture to the study of multi-band systems.
As demonstrated in Ref.~\cite{ammendola2025highdimensionalprogrammablephotoniccircuits}, our platform can also be used at the single- and multi-photon domain, allowing, in the future, to test QW processes of spatially correlated or anti-correlated quantum states \cite{paneru2024nonlocal}.

\section{Methods}
\subsection{Experimental setup}
The experiment was realized by feeding the setup with an 808~nm diode laser coupled to a single-mode fiber in order to prepare a Gaussian beam as an input to the 3-SLM stack. The detailed setup is reported in the Supplementary Figure S2-a. The beam was magnified to have $\sim 1$ cm beam waist and prepared with horizontal polarization. The phase masks implementing the QW were displayed on three reflective Hamamatsu LCOS-SLMSs. SLM1 and SLM2 have 1272$\times$1024 square pixels, each 12.5 $\mu$m wide, while SLM3 has a lower resolution (792$\times$600 pixels, each 20 $\mu$m wide). A $4f$ imaging system was built between each pair of SLMs to minimize propagation effects between one mask and the next. A pinhole was added to the first $4f$ imaging to select the first diffraction order of SLM1, which is needed to initialize the input state into an array of beamlets, which effectively implements the discrete sampling of the BZ (see the next section). The reflection angle at each SLM was less than $10^{\circ}$ to minimize reflection-dependent polarization effects, which were additionally compensated by calibrating the color scale of the phase masks to yield the desired optical retardation. The lattice space distributions were recorded via a CMOS camera placed at the focal plane of a Fourier-transforming lens mounted after SLM3. To implement simulations of wavepackets, a pinhole with variable aperture was placed before SLM1, allowing one to select the number of reciprocal lattice sites occupied by the initial state, as well as the initial average momentum of the wavepacket, which could be chosen by translating the pinhole along the transverse plane.

\subsection{Design of the phase masks and experimental imperfections}
To find the desired phase masks, we used the method demonstrated in Ref.~\cite{ammendola2025highdimensionalprogrammablephotoniccircuits}, which we briefly review here. The aim is to find the values of the retardation parameters $\Delta_i(x_h,y_h)$, with $i=1,2,3$ and $(x_h,y_h)$ labeling the pixel coordinates of the SLM planes, for the operator \begin{align}\mathcal{S}(x_i,y_i)=&e^{i\frac{\Delta_0}{2}}S_{0}(\Delta_3)\cdot HWP(-22.5\degree)\cdot \cr& S_{0}(\Delta_2)\cdot HWP(22.5\degree)\cdot S_{0}(\Delta_1),\end{align} with ${\Delta_0=\sum_i \Delta_i/2}$ and 
\begin{equation}
{S_0(\Delta_i)=\cos(\Delta_i/2)\mathbb I+i\sin(\Delta_i/2)(\ket{R}\bra{L}+h.c)}    
\end{equation}
is the SLM operator in the circular polarization basis, except for a global phase shift that is accounted for in $\Delta_0$~\cite{ammendola2025highdimensionalprogrammablephotoniccircuits}. $\mathcal{S}(x_h,y_h)$ should be equal to the $t$-step unitary $\mathcal{U}(q_{x,h},q_{y,h})$, where the momentum space coordinates $(q_{x,h},q_{y,h})$ is also to be mapped in the SLM plane coordinates $(x_h,y_h)$. Different sets of analytical solutions to the problem can be found by decomposing both $\mathcal{S}$ and $\mathcal{U}$ in Pauli matrices and equating the coefficients. 
Defining ${\alpha=\Delta_2/2}$, ${\beta=\frac{\Delta_1+\Delta_3}{2}}$, and ${\gamma=\frac{\Delta_1-\Delta_3}{2}}$, we can express a possible set of solutions as 
\begin{equation}
\begin{cases}
\alpha_1=\text{atan2}\left(\alpha_x,\alpha_y\right)\\
\beta_1=\text{atan2}\left(\beta_x,\beta_y\right)\\
\gamma_1=\text{atan2}\left(\gamma_x,\gamma_y\right)
\end{cases}
\label{eqn:solutions}
\end{equation}
where atan2(${x,y}$) is the two-argument arc-tangent function, and
\begin{equation}
\begin{split}
\alpha_x&=\cos\theta\sin E/\sqrt{1-h},\\
\alpha_y&=-\sin E \sin\theta\sin\phi/\sqrt{1-h},\\
\beta_x&=\sqrt{1-h},\\
\beta_y&=-\sqrt{h},\\
\gamma_x&=-\cos\phi \sin E \sin\theta/\sqrt{h},\\
\gamma_y&=-\cos E/\sqrt{h},
\end{split}
\end{equation}
where $h=\cos{E}^2-\sin{E}^2\sin{\theta}^2\cos{\phi}^2$, $E$ is the quasi-energy, and $\theta$ and $\phi$ are, respectively, the polar and azimuthal angle in the unit sphere representation of the Bloch eigenvector $\mathbf{n}$.
These solutions allow to extract $\Delta_1$, $\Delta_2$, and $\Delta_3$ defined  Mod$(4\pi)$, which are adequately wrapped to yield phase masks defined Mod$(2\pi)$. More details can be found in Ref.~\cite{ammendola2025highdimensionalprogrammablephotoniccircuits}. 

In this work, holograms were calculated only for those values of the BZ that are allowed by the periodic boundary conditions. Moreover, for a proper implementation of a cyclic QW, the input beam should not be a smooth Gaussian, since it would populate regions of the reciprocal lattice that are not allowed in the cyclic process. It is thus more appropriate to use an input mode made of an array of Gaussian spots with narrow width (which, ideally, should be of the order of one SLM pixel). The transverse amplitude of the input field is of the form
\begin{align}A(\textbf{q})=\sum_{\textbf{h}} c_{\textbf{h}}\,g_\sigma(\textbf{q}-\textbf{q}_{\textbf{h}})    \label{eq:spots}\end{align}
where $c_{\textbf{h}}$ are complex coefficients, and the functions $g_\sigma$ should have a characteristic width $\sigma$ narrower than the reciprocal lattice spacing. Ideally, they should be Dirac-delta functions. Small experimental deviations from the theoretical model are expected due to the finite width of $g_\sigma$. In our experiment, we effectively encode this input state by superimposing an additional mask on the first SLM to the one corresponding to the QW unitary computed from Eq.~\eqref{eqn:solutions}. Figure~S2-b,c shows examples of the phase masks used in the experiment for the 1D and 2D protocols, respectively. The insets show how the first SLM is used to encode both the first mask $\Delta_1$ corresponding to QW $\hat{U}^t$ and to shape the input Gaussian beam (having a waist comparable with the SLM aperture) into an array of narrower light spots. A diffraction grating is applied only in the areas where we want the light to be imaged on the next SLM, therefore a pinhole (Ph2) is placed after the first SLM to select the first diffraction order, thus realizing the desired mask. This approach was used to minimize the number of SLMs needed for the setups, however, it imposes some limits on the quality of the 2D quantum walk results due to the low resolution of the SLMs and introduces losses in the setup. These limits can be surpassed, in principle, with an additional device (e.g., a fourth SLM) that prepares the mode in Eq.~\eqref{eq:spots}. These modes could be accurately realized using, for instance, micro-lens arrays. 
More quantitatively, the calculated mask for the first SLM performs the action 
\begin{align*}
SLM_1(x,y)=&\sum_{\mathbf{h}}e^{-i(\Delta_0(x-x_{\mathbf{h}},y-y_{\mathbf{h}})-\Delta_1(x-x_{\mathbf{h}},y-y_{\mathbf{h}}))}\cr&\times c_{\mathbf{h}}g_{\sigma}(x-x_{\mathbf{h}},y-y_{\mathbf{h}}),
\label{eq:slm1relmask}
\end{align*}
where $c_{\mathbf{h}}$ and $g_{\sigma}$ are the same as defined in Eq.~\eqref{eq:spots}. The corresponding quantum state at the output of the quantum walk setup will be given by 
\begin{align}
\ket{\psi}=\int_{\text{BZ}} d^Dq \sum_{\mathbf{h}}c_{\mathbf{h}}g_{\sigma}(\mathbf{q}-\mathbf{q}_{\mathbf{h}}) (\mathcal{U}(\mathbf{q}_{\mathbf{h}})\ket{H})\otimes\ket{\mathbf{q}}.
\end{align}
Note that, in the limit $g_\sigma(\mathbf{q}-\mathbf{q}_{\mathbf{h}})\rightarrow \delta(\mathbf{q}-\mathbf{q}_{\mathbf{h}})$, we recover Eq.~\eqref{eq:QWstate}. It is useful to consider the state representation in the lattice position basis, i.e., to calculate the wavefunction $\psi(\mathbf{m},t)$:
\begin{align}
\psi(\mathbf{m},t)=\int_{\text{BZ}} d^Dq \sum_{\mathbf{h}}c_{\mathbf{h}}g_{\sigma}(\mathbf{q}-\mathbf{q}_{\mathbf{h}}) e^{i\mathbf{m}\cdot\mathbf{q}}\mathcal{U}(\mathbf{q}_{\mathbf{h}})\ket{H}.
\end{align}
Multiplying and dividing by $e^{i\mathbf{m}\cdot\mathbf{q}_{\mathbf{h}}}$, and defining the Fourier transform of $g_{\sigma}$: $\tilde{g}_{\sigma}(\mathbf{m})=\int g_{\sigma}(\mathbf{q})\exp(i\mathbf{m}\cdot \mathbf{q})d^Dq$, we obtain
\begin{align}
\psi(\mathbf{m},t)= \tilde{g}_{\sigma}(\mathbf{m})\sum_{\mathbf{h}}c_{\mathbf{h}}\mathcal{U}(\mathbf{q}_{\mathbf{h}})\ket{H}.
\end{align}
This shows that the state obtained from the experimental setup differs from that of an ideal cyclic QW by the multiplicative factor $\tilde{g}(\mathbf{m})$ which can be accounted for, either from calibration measurements or by preparing functions $g_{\sigma}$ narrow enough to be essentially flat in the region of interest given by $N_x\times N_{y}$ lattice sites, as was done in this work. The envelope $\tilde{g}_{\sigma}(\mathbf{m})$ determines the relative intensities of the additional replicas of the QW distribution, such as those shown in Fig.~\ref{fig:simplified setup}-b. Note that, in the limit $g_\sigma(\mathbf{q}-\mathbf{q}_{\mathbf{h}})\rightarrow \delta(\mathbf{q}-\mathbf{q}_{\mathbf{h}})$,  $\tilde{g}_{\sigma}$ is a uniform distribution and one should observe, in the far field of the SLMs, infinite identical copies of the cyclic QW. We also point out a tradeoff that must be considered in our implementation of the state of Eq.~\eqref{eq:spots} via Eq.~\eqref{eq:slm1relmask}: if $\sigma$ is too small and/or $\mathbf{N}$ is too large, effects from the zero diffraction order may overlap with the first order introducing experimental deviations from the theory; if $\sigma$ is too large, the envelope function $\tilde{g}_{\sigma}$ can affect significantly the probability distribution, as we particularly observed in the case of the QW on the torus. In the future, this could be circumvented by generating the beam array with micro-lenses or an additional phase-only SLM, which will allow for the implementation of cyclic walks with higher fidelity on a large scale. 
 \subsection{Details on the QW protocols $\hat{U}_{1D}$ and $\hat{U}_{2D}$}
 Here, we report explicit expressions for the quasi-energy band dispersion of the implemented QW protocols. These expressions can be found by analytically decomposing the corresponding single-step unitary operators in the quasi-momentum space and the basis of Pauli matrices. For $\hat{U}_{1D}(\delta=\pi)$, the energy dispersion is 
 \begin{align}
 E_{1D}(q)=\arctan(\cos(q), \sqrt{2 \sin^2(q)+\cos^2(q)}),
 \end{align}
 while, for $\hat{U}_{2D}(\delta=\pi/2)$, we have
 \begin{align}
 E_{2D}(q_x,q_y)&=\arccos\big(1-\cos(q_x)-\cr&\cos(qx-q_y)-\cos(q_y))/2\sqrt{2}
\big). \end{align}
Additional details on these protocols are reported in Refs.~\cite{cardano2015quantum,cardano2017detection, DErrico2020}.

\subsection{Recurrent and non-recurrent dynamics in QWs on 3-cycles}

The experiments with ${D=1}$ and ${N=3}$ shown in Fig.~\ref{fig:locstates}-a were realized implementing variations of the $\hat{U}_{1D}$ protocol, where $\hat{W}$ was either replaced with $\sigma_0, i\sigma_1$, or left unchanged: ${\hat{W}=(\sigma_0+i\sigma_2)/\sqrt{2}}$. The following relations hold for the quasi-energies of these protocols: ${\{E(2\pi/3)=E(4\pi/3)=E(2\pi)=\pi/2\}}$ for ${\hat{W}\rightarrow \sigma_0}$, ${\{E(2\pi/3)/E(4\pi/3)=1,E(4\pi/3)/E(2\pi)=2/3\}}$ for ${\hat{W}\rightarrow\sigma_1}$, and $\{E(2\pi/3)/E(4\pi/3)=1,E(4\pi/3)/E(2\pi)=0.513298\}$ for $\hat{W}\rightarrow(\sigma_0+i\sigma_2)/\sqrt{2}$, thus having, respectively, $t^*=2,3$, and $t^*>30$. Almost perfect recurrence is expected for $t=253$ in the latter case. The 30-step dynamics obtained experimentally is shown in Fig.~\ref{fig:locstates}-a. Note that, strictly speaking, the recurrence time for the first case, $\hat{W}\rightarrow\hat{\sigma}_0$, should be $t^*=4$, however, for $t=2$, the process is equal to $-\mathbb I$.

\subsection{Propagation of scalar wavepackets with zero group velocity in cyclic lattices}
\label{sec:methodsE}
Here, we develop a simplified model to gain a better qualitative understanding of the results in Fig.~\ref{fig:1dwavepackets2}-b,c. Consider a Gaussian wavepacket in the form $\ket{\psi_0}\propto\sum_h e^{-(q_h-q_0)^2w^2/4}\ket{q_h}\otimes\ket{\mathbf{n}_+(q_0)}$. We will assume that the width $2/w$ of the wavepacket in the reciprocal lattice is small enough so that ${\ket{\mathbf{n}_+(q_0)}\approx \ket{\mathbf{n}_+(q_h)}}$ when $|q_h-q_0|\lesssim 2/w$ (scalar approximation). Moreover, we focus on the special case where $v_g(q_0)=0$. Within these assumptions, the evolved state is $\ket{\psi_t}\propto\sum_h e^{-(q_h-q_0)^2w^2/4}e^{-iE(q_h)t}\ket{q_h}\otimes\ket{\mathbf{n}_+(q_0)}$.
The probability of finding the particle in the position $m$ at the time step $t$ is
\begin{align}
P(m,t)=&\abs{\braket{m}{\psi_t}}^2\cr=&\abs{\sum_h e^{i q_h m}e^{-(q_h-q_0)^2\frac{w^2}{4}}e^{iE(q_h)t}}^2.
\label{eq:Pwp2}
\end{align}
The evaluation of this expression is in good agreement with the full QW simulations (see Supplementary Figure S1), showing how the scalar approximation holds for the considered wavepacket.

As outlined in the main text, an intuitive understanding of the observed breathing behavior can be gained by considering a minimal amount of terms in the sum of Eq.~\eqref{eq:Pwp2}. In the following, we consider the cases where $q_0=q_h$, for some index $h$, and $q_0=q_h\pm\pi/N$. 

\textit{Interference of 3 plane waves ($q_0=q_h$).}

If $q_0=q_h$, for some $h$, and $E(q_h)$ is a local maximum or minimum of the energy band, then one can consider 3 plane wave contributions with ${q_0=q_h}$, and $q_{h\pm1}$. This case is equivalent to ${q_0=\pi}$ in the main text. Then, assuming ${E(q_{h+1})=E(q_{h-1})}$,  $P(m,t)\approx |e^{i(E(q_h)t-q_h m)}+2e^{-(2\pi/N)^2w^2/4}e^{i E(q_{h+1})t}\cos(m 4\pi/N)|^2$. This probability distribution exhibits an oscillating behavior both in $m=0$ and the antipodal point $m\approx N/2$, with a period dependent on the energy difference $|E(q_h)-E(q_{h-1})|$. The relative phase $e^{iq_h m}$ dictates the time steps where maxima and minima emerge for a fixed $m$. 

\textit{Interference of 4 plane waves ($q_0=q_h-\pi/N$).}

This case is equivalent to $q_0=0$ in the main text.
If ${q_0=q_h-\pi/N}$, we can consider expanding $P(m,t)$ in terms of the 4 dominant eigenstates (considering the first 2 is not sufficient to reproduce the expected breathing dynamics). If $q_0$ corresponds to a maximum or minimum of the energy band, then we can again assume $E(q_0+\pi/N)=E(q_0-\pi/N)$ and $E(q_0+3\pi/N)=E(q_0-3\pi/N)$. Hence, we have $P(m,t)\approx 4|e^{-(\pi/N)^2w^2/4}e^{iE(q_0-\pi/N)}\cos(m\pi/N)+e^{-(3\pi/N)^2w^2/4}e^{iE(q_0-3\pi/N)}\cos(3m\pi/N)|^2$. While for ${m\approx 0}$, $P(m,t)$ exhibits an oscillating behavior with characteristic period $|E(q_0-\pi/N)-E(q_0-3\pi/N)|^{-1}$, at the antipodal site, $m\approx N/2$, we find $P(m,t)\approx 0$.

\bibliography{bibliography.bib}

\subsection*{Funding}
This work was supported by the Canada Research Chairs (CRC), and Quantum Enhanced Sensing, Imaging (QuEnSI) Alliance Consortia Quantum grant, and PNRR MUR project PE0000023-NQSTI. MGA further acknowledges support from MITACS and the Italy-Canada Innovation Award.
\subsection*{Author contributions} A.D. conceived the idea. F.D.C. and A.D., with contributions from F.G. and E.K., developed the theory. F.D.C., with contributions from M.G.A. and A.D. devised the phase masks. M.G.A. and N.D., prepared the experimental setup. A.D., with contributions from N.D., and R.G. collected and analyzed the data. L.S. developed the code for automated data acquisition. F.C and E.K supervised the project. A.D. prepared the first draft of the manuscript. All authors contributed to the revision of the manuscript.
\subsection*{Competing interest}
The authors declare that they have no competing interests.
\subsection*{Data and materials availability}
All data are available in the main text or the supplementary materials.

\clearpage
\onecolumngrid
\renewcommand{\figurename}{\textbf{Figure}}
\setcounter{figure}{0} \renewcommand{\thefigure}{\textbf{S{\arabic{figure}}}}
\setcounter{table}{0} \renewcommand{\thetable}{S\arabic{table}}
\setcounter{section}{0} \renewcommand{\thesection}{S\arabic{section}}
\setcounter{equation}{0} \renewcommand{\theequation}{S\arabic{equation}}
\onecolumngrid
\begin{center}
{\Large Supplementary Material for: \\ Programmable photonic quantum walks on lattices with cyclic, toroidal, and cylindrical topology}
\end{center}
\vspace{1 EM}
\section{Propagation of scalar wavepackets with zero group velocity in cyclic lattices: parabolic energy band approximation}
\begin{figure}[h!]
\includegraphics[width=\columnwidth]{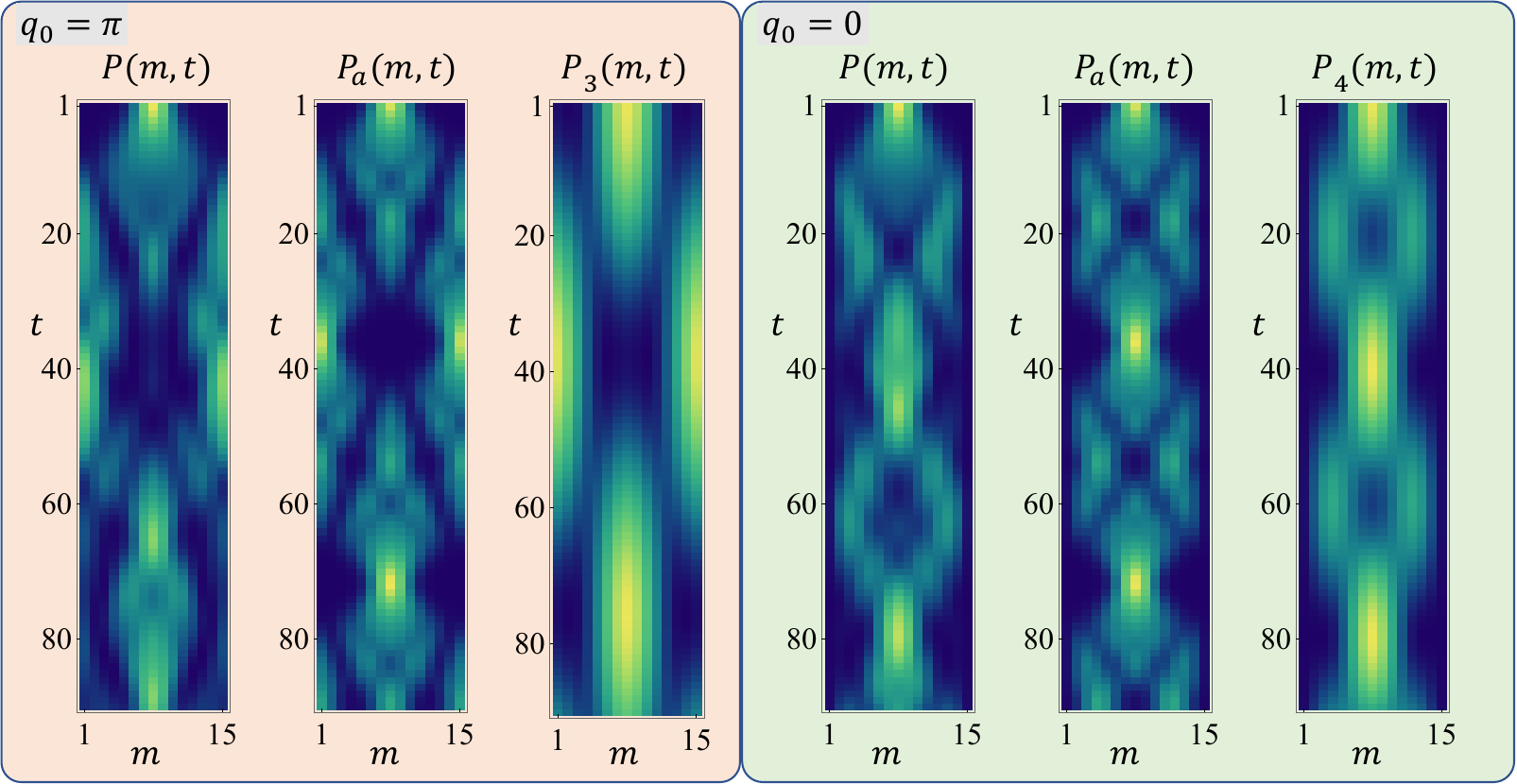}
    \caption{\textbf{Breathing dynamics of scalar wavepackets with $v_g=0$.} $P(m,t)$ corresponds to Eq.~\eqref{eq:scalarwp}. $P_a(m,t)$ is evaluated from Eq.~\eqref{eq:scalarGVD} expanding the energy dispersion to second order. $P_{3/4}(m,t)$ are obtained by considering only the 3 (4) values of $q_h$ closest to $q_0$ in Eq.~\eqref{eq:scalarwp} and illustrate the trend calculated in the Methods (Sec. V E).}
    \label{fig:GVD}
\end{figure}

As shown in the Methods, the probability of finding a particle in the lattice site $m$ at the time step $t$ is
\begin{align}
P(m,t)=&\abs{\braket{m}{\psi_t}}^2\cr=&\abs{\sum_h e^{i q_h m}e^{-(q_h-q_0)^2\frac{w^2}{4}}e^{iE(q_h)t}}^2.
\label{eq:scalarwp}
\end{align}
It is worth considering a further approximation, which consists in evaluating $P(m,t)$ using the expansion $E(q_h)=E(q_0)+1/2 (d^2E/dq^2)|_{q_0}(q_h-q_0)^2\equiv E_0+\beta(q_h-q_0)^2$, where the first-order term was neglected assuming zero group velocity.
It follows that
\begin{align}
P(m,t)\approx P_a(m,t)=\abs{\sum_h e^{i q_h m}e^{-(q_h-q_0)^2\left[\frac{w^2}{4}+i\beta t\right]}}^2.
\label{eq:scalarGVD}
\end{align}
This expression describes the wavepacket in terms of the group velocity dispersion $\beta$, the wavepacket's width $w$, and the sampling of the Brillouin zone $q_h$. We consider the cases $q_h(s)-q_0=2\pi h/N-s\pi/N$ with $s=0,1$ (corresponding, respectively, to $q_0=\pi$ and $q_0=0$ in our experiment). Figure~\ref{fig:GVD} shows the trend of $P(m,t)$ compared with the second-order approximation $P_a(m,t)$ for our protocol. We can see that $P_a(m,t)$ captures the main qualitative effects, i.e., that for $s=1$ the wavepacket periodically tunnels between opposite positions in the lattice space, while for $s=0$ it periodically splits in two lobes then refocusing to the original position. However, the second-order expansion in $P_a$ fails to predict the right refocusing time and finer details in the probability distribution, although the agreement can be improved by fine-tuning the parameters $\beta$ and $w$. The full expression of $P$ is, instead, in good agreement with the experiment, as well as with a full QW simulation, except for the emergence of fringe patterns, which would require including the polarization degree of freedom in the description and, in particular, the input polarization state. We also show the evaluation of $P(m,t)$ for a minimal number of terms contributing to the sum in Eq.~\eqref{eq:scalarwp}. As shown in the Methods, the basic features of the breathing dynamics are captured considering only 3 eigenstates for $q_0=\pi$ and 4 eigenstates for $q_0=0$.

\section{Connection between the wavefunction on the infinite and cyclic lattice.}

Here, we show that the wavefunction calculated for the infinite lattice case can be used to derive the expression of the wavefunction on the cyclic lattice. To retrieve the latter, we prove that it is sufficient to coherently add the contributions of all the lattice sites spaced by $N$.
To model a quantum walk constrained to a ring consisting of $N$ discrete lattice sites, we restrict the allowed momenta to a finite set of discrete Bloch momenta ${q_n = 2\pi n/N}$, where $n = 0, 1, \dots, N-1$. This momentum-space discretizations is enforced by the projection operator $P_N = \sum_{n=0}^{N-1} |q_n\rangle \langle q_n|$. Applying $P_N$ to the infinite-lattice state yields the finite-lattice state at time $t$: $|\psi_N(t)\rangle = P_N |\psi_\infty(t)\rangle=\sum_{n=0}^{N-1} \phi_w(q_n) |q_n\rangle \otimes U_{q_n}^t |\phi_c\rangle$, where $\phi_w(q_n)$ is the momentum wavefunction.

In the discrete position basis defined by orthonormal states $\{|x_m\rangle\}_{m=0}^{N-1}$, where the overlap with momentum states is given by $\langle x_m|q_n\rangle = e^{iq_n x_m}/\sqrt{N}$, the matrix elements of the projection operator become:
$\langle x_m | P_N | x_{m'} \rangle = \sum_{r \in \mathbb{Z}} \delta_{m', m + rN}$, with $r$ an integer. Consequently, the finite-lattice wavefunction in the discrete position basis is

\begin{equation}\label{psim}
\psi_{N}(x_m, t)=\sum_{r\in\mathbb{Z}}\psi_\infty(x_m+rN,t). 
\end{equation}

Note that the summation enforces the periodic boundary condition $\psi_N(x+N)=\psi_N(x)$. According to Eq.~\eqref{psim}, the cyclic wavepacket is obtained by taking the free-space wavefunction $\psi_\infty(x,t)$ on the infinite line and wrapping it around the ring by adding up all of its images shifted by every integer multiple of $N$. This approach allows one to obtain an analytical expression of the probability distribution of the cycle when the energy dispersion can be expanded to second order in $q$.

\textit{Gaussian State.} As an application, we derive the closed-form expression for the evolution of Gaussian wavepackets. We focus on an experimentally relevant case where the walker is initialized in a Gaussian state \(\phi_w(q)=\frac{1}{(2\pi\sigma_q^{2})^{1/4}}\exp\bigl[-\tfrac{(q-q_0)^2}{4\sigma_q^{2}}\bigr]\). As in the Methods section, we assume a constant polarization and smoothly varying eigenstates. Hence, we can neglect the \qo{coin} terms. Under arbitrary dispersion $E(k)$, the evolved state on the infinite lattice is 

\begin{equation}
\psi_\infty(x,t)
=\frac{1}{\sqrt{2\pi}}\int_{-\pi}^{\pi}
dq\phi_w(q)e^{iqx -i t E(q)}. 
\label{eq:scalarpsit}
\end{equation}

Inserting the integral form of $\psi_\infty$ in Eq.~\eqref{psim} yields

$$
\psi_{N}(x,t)
=\frac{1}{\sqrt{2\pi}}
\sum_{r\in\mathbb Z}
\int_{-\pi}^{\pi}dq
\phi_w(q)\,
e^{iq(x+rN)-it E(q)}.
$$

Swapping sum and integral operators, and applying the identity $\sum_{r\in\mathbb Z}e^{\,i q rN}
=\frac{2\pi}{N}\sum_{m\in\mathbb Z}\delta\!\Bigl(q-\tfrac{2\pi m}{N}\Bigr)$, we obtain

$$
\psi_{N}(x,t)
=\frac{1}{\sqrt{2\pi}}
\int dq\;\phi_w(q)e^{iqx-t E(q)t}\;
\frac{2\pi}{N}\sum_{m\in\mathbb Z}\delta\!\bigl(q-q_m\bigr).
$$

Performing the $\delta$-integral gives

$$
\psi_{N}(x,t)
=\frac{1}{\sqrt{N}}
\sum_{m=-\infty}^{\infty}
\phi_w\bigl(q_m\bigr)e^{iq_mx -it E(q_m)}. 
$$

From this, we can also obtain the on-ring probability density $P_N(x,t)=|\psi_N(x,t)|^2$:

$$
\begin{aligned}
P_N(x,t)
&=\sum_{r\in\mathbb Z}
C_r(t)\;e^{\,i\,\tfrac{2\pi r}{N}\,x},
\end{aligned}
$$
where we have defined the correlation coefficients
$$
C_r(t)
=\frac{1}{N}
\sum_{n\in\mathbb Z}
\phi_w(q_{n+r})\,\phi_w^*(q_n)e^{-it[E(q_{n+r})-E(q_n)]},
$$
with $C_{-r}(t)=C_r^*(t)$ ensuring $P_N(x,t)$ is real. For a Gaussian wavepacket, the correlation function assumes the following form:
$$
C_r(t)
=e^{-\tfrac{(r\,\Delta q)^2}{4\sigma_q^2}}f_r,
$$
with $f_r=\frac{1}{N\sqrt{2\pi\sigma_q^2}}
\sum_{n\in\mathbb Z}
\exp\!\Bigl[
-\frac{(q_n-q_0)^2}{2\sigma_q^2}
-\frac{r\,\Delta q\,(q_n-q_0)}{2\sigma_q^2}
-it\bigl(E(q_{n+r})-E(q_n)\bigr)
\Bigr]$. The factor $\propto e^{-(r\,\Delta q)^2/(4\sigma_q^2)}$ suppresses large-$|r|$ harmonics. 

Assume now the energy dispersion is smooth enough to be Taylor-expanded to the second order as \(E(q) \approx E(q_0) + v(q - q_0) + \frac{D}{2}(q - q_0)^2\), where \(v_g := E'(q_0)\) is the group velocity, and \(D := E''(q_0)\) is the dispersion coefficient that quantifies curvature in the energy spectrum. Starting from Eq.~\eqref{eq:scalarpsit} and performing a straightforward Gaussian integration, we obtain

\begin{equation}\label{psiG}
\psi_{\infty}(x,t)= \langle x|\psi(t)\rangle=\mathcal{N}(t) e^{i(q_0 x - t E_0)}\exp\left[-\frac{(x - v_gt)^2}{4A(t)}\right],
\end{equation}
where $\mathcal{N}(t)=\frac{1}{(2\pi\sigma_q^{2})^{1/4}}\sqrt{\frac{\pi}{A(t)}}$ is a normalization factor and \(A(t)=\frac{1}{4\sigma_q^2}+i\frac{tD}{2}\).  Plugging Eq.~\eqref{psiG} into Eq.~\eqref{psim}, we obtain the closed-form wavefunction for the cyclic walk:

\begin{equation}\label{psiN}
\psi_{N}(x_m,t)=\frac{1}{\sqrt{\mathcal S(t)}}\psi_\infty(x_m,t)\Theta_3(z(t),\gamma(t)),
\end{equation}
where $\Theta_{3}(z,\gamma)=\sum_{\ell=-\infty}^{\infty}\gamma^{\ell^{2}}e^{2i\ell z}$ is the Jacobi Theta function, with $z(t)=\;\frac{q_0\,N}{2}\;+\;i\,\frac{(x_m-v_g t)\,N}{4A(t)}$ and $\gamma(t)=\exp\Bigl[-\tfrac{N^2}{4A(t)}\Bigr]$, and normalization  $\mathcal S(t)=\frac{N}{\sqrt{2\pi}\,\sigma_x(t)}\Theta_{3}\!\bigl(\tfrac{q_0N}{2}, |\gamma(t)|\bigr)$.

As an immediate consequence of Eq.~\eqref{psiN}, one can find the probability distribution and fidelity for the cyclic walk as 

\[
P_N(x_m,t)=\frac{1}{\mathcal S(t)}P_\infty(x_m,t)|\Theta(z(t),q(t))|^{2},  
\]

with 
\[
P_{\infty}(x,t) = \frac{1}{\sqrt{2\pi \sigma_x^2(t)}} \exp\left[-\frac{(x - v_gt)^2}{2\sigma_x^2(t)}\right].
\]

\section{Supplementary Figures}

Figure S1 shows the 1D breathing dynamics at $v_g=0$ under different approximations. From left to right insets: scalar approximation, second-order expansion of the energy dispersion, and 3 or 4-wave expansion.

Figure S2 shows a detailed experimental setup.

Figure S3 shows all the results from the QW on the torus, compared with theory.

Figure S4 shows all the results from the QW on the cylinder, compared with theory.
\begin{figure*}[h!]
\includegraphics[width=0.9\textwidth]{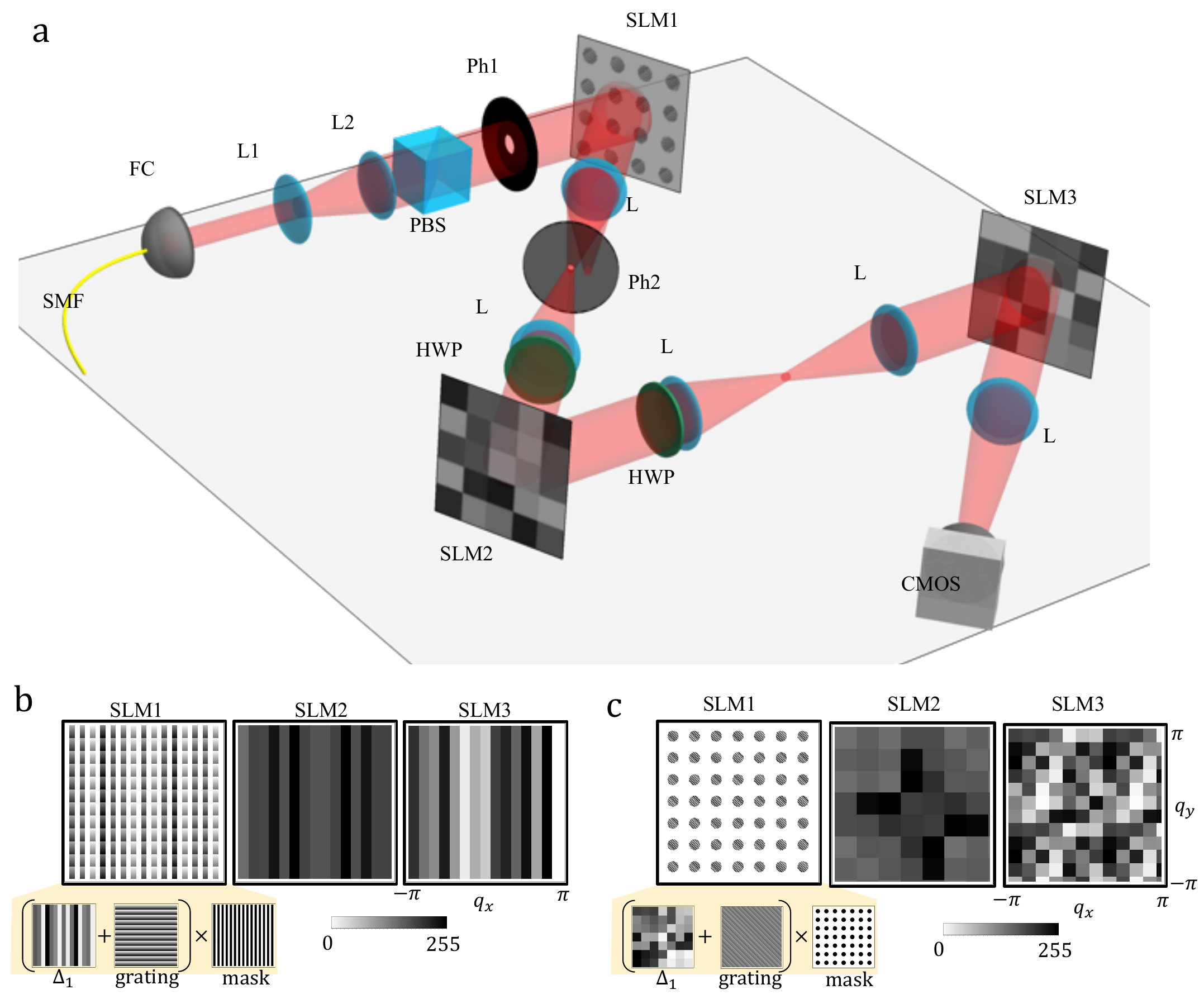}
\caption{\textbf{Detailed experimental setup.} SMF: single-mode fiber. FC: Fiber output coupler. L1 and L2: beam magnification stage. PBS: Polarizing beam splitter. Ph1: pinhole that allows choosing the width $2/w$ of the input Gaussian states. L: Lens. Ph2: pinhole selecting the first diffraction order from SLM1. HWP: half-wave plates rotated at $\pm$22.5° with respect to the extraordinary axis of SLM1. CMOS: camera used to measure far-field intensity distributions. b.-c. Examples of phase masks applied on each SLM. The SLM1, combined with Ph2, simultaneously encodes part of the unitary process and the structuring of the input state into an array of spots using the combination of masks illustrated in the inset.}
    \label{fig:ST2}
\end{figure*}
\begin{figure*}
\includegraphics[width=\textwidth]{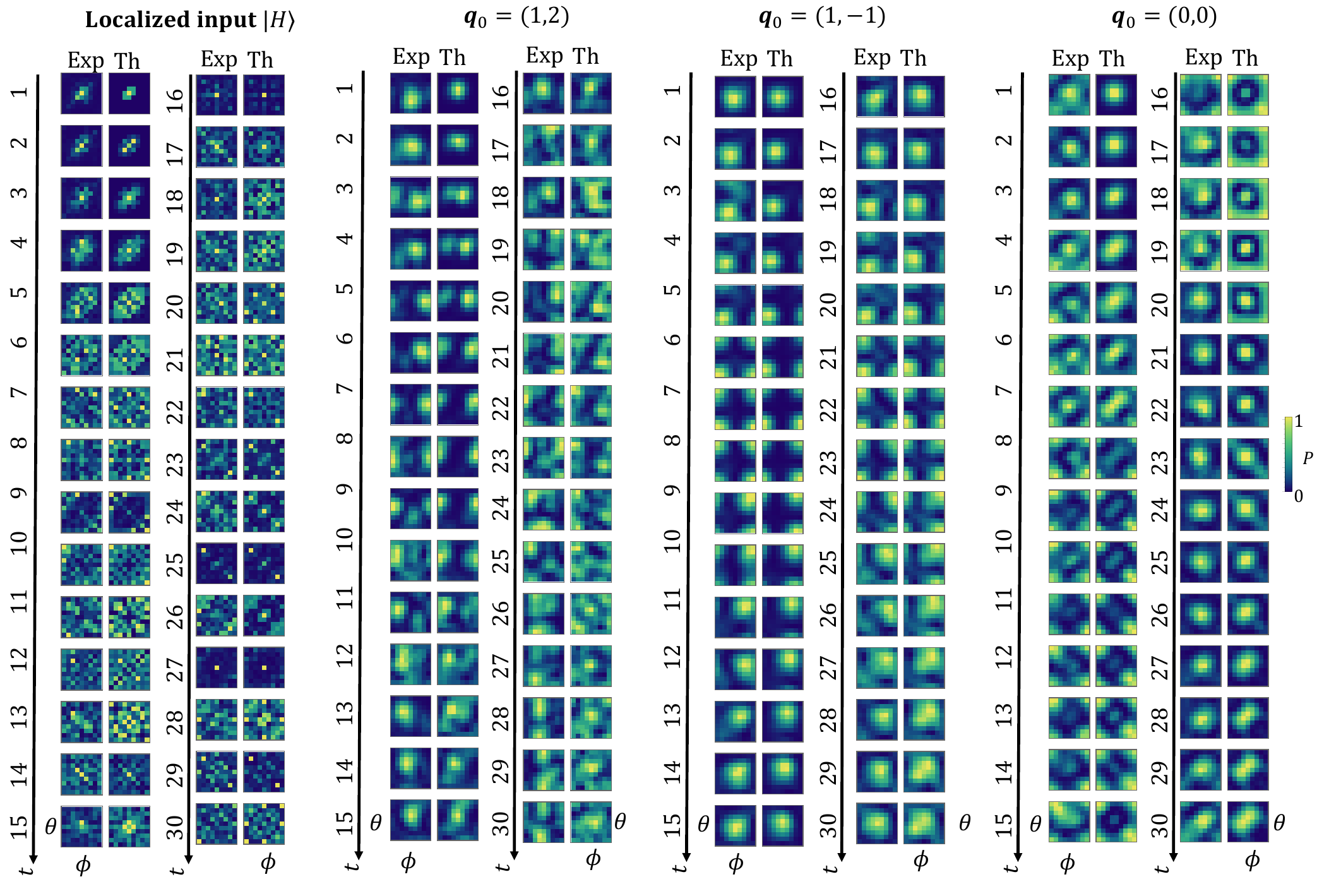}
    \caption{\textbf{QWs on a torus for initially localized states and wavepackets.} Experimental and theoretical QW distributions on a torus with 9$\times$9 lattice sites over 30 time steps. Initial states are all horizontally polarized and prepared either as localized states or as wavepackets. The theory assumes a Gaussian wavepacket. }
    \label{fig:ST2}
\end{figure*}

\begin{figure*}
\includegraphics[width=\textwidth]{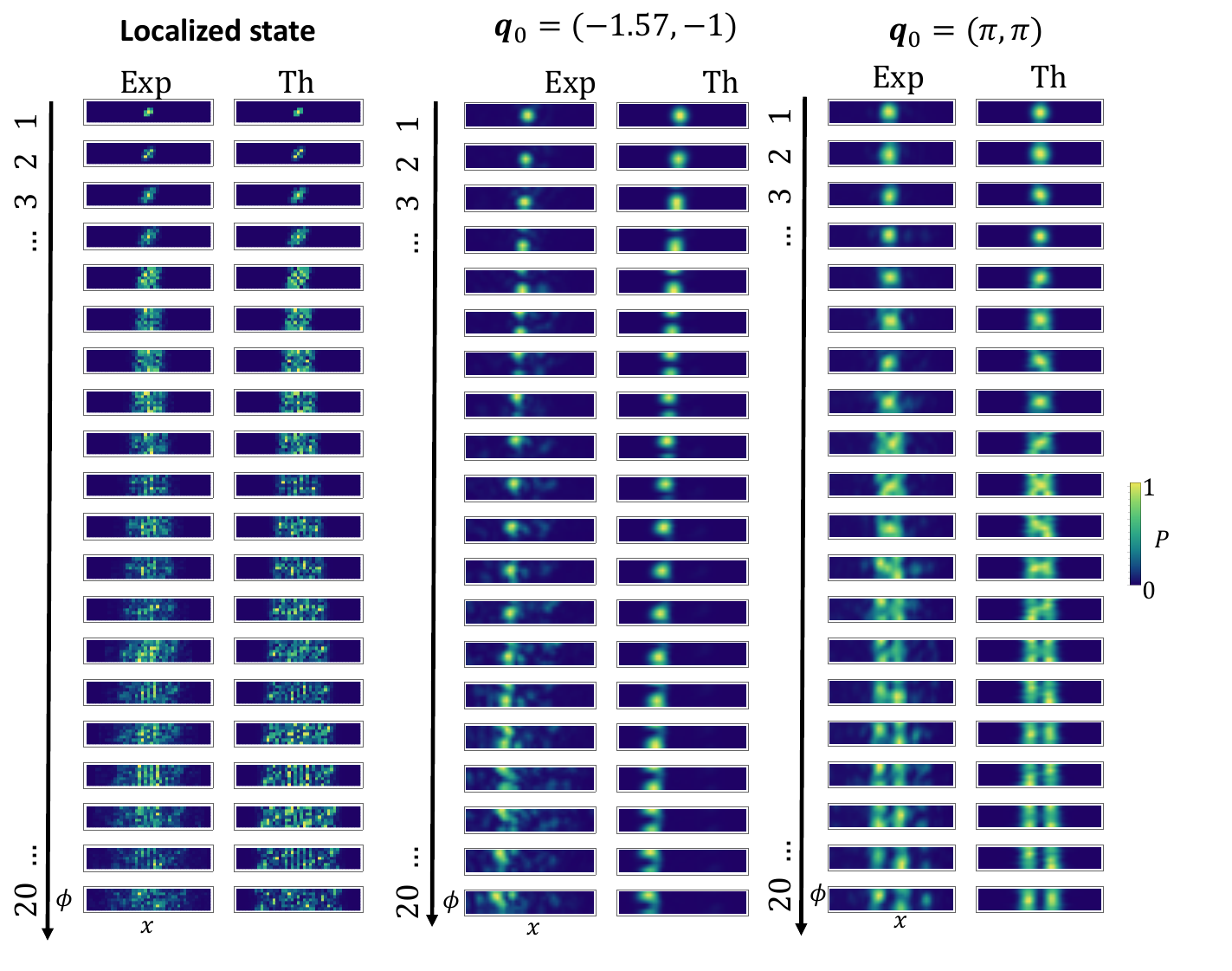}
    \caption{\textbf{QWs on a cylinder for initially localized states and wavepackets.} Experimental and theoretical QW distributions on a cylinder, with 7 lattice sites in the cyclic direction over 30 time steps. Initial states are all horizontally polarized and prepared either as localized states or as wavepackets. The theory assumes a Gaussian wavepacket.}
    \label{fig:ST2}
\end{figure*}








\end{document}